\documentclass[preprint, superscriptaddress, showpacs,preprintnumbers,amsmath,amssymb,prb,endfloats*]{revtex4}
\usepackage{graphicx}

\begin{document}
\pdfoutput=1
\thispagestyle{empty}

\title{Modifying the Casimir force between indium tin oxide film
and Au sphere}

\author{ A.~A.~Banishev}
\affiliation{Department of Physics and
Astronomy, University of California, Riverside, California 92521,
USA}

\author{C.-C.~Chang}
\affiliation{Department of Physics and
Astronomy, University of California, Riverside, California 92521,
USA}

\author{R.~Castillo-Garza}
\affiliation{Department of Physics and
Astronomy, University of California, Riverside, California 92521,
USA}

\author{G.~L.~Klimchitskaya}
\affiliation{North-West Technical University, Millionnaya Street 5,
St.Petersburg, 191065, Russia}

\author{V.~M.~Mostepanenko}
\affiliation{Noncommercial Partnership ``Scientific Instruments'',
Tverskaya Street 11, Moscow, 103905, Russia}

\author{U.~Mohideen}
\affiliation{Department of Physics and
Astronomy, University of California, Riverside, California 92521,
USA}

\begin{abstract}
We present complete results of the experiment on measuring the
Casimir force between an Au-coated sphere and an untreated or,
alternatively, UV-treated indium tin oxide film deposited on
a quartz substrate. Measurements were performed using an atomic
force microscope in a high vacuum chamber. The measurement
system was calibrated electrostatically. Special analysis of
the systematic deviations is performed, and respective
corrections in the calibration parameters are introduced.
The corrected parameters are free from anomalies discussed in the
literature. The experimental data for the Casimir force from two
measurement sets for both untreated and UV-treated samples are
presented. The random, systematic and total experimental errors are
determined at a 95\% confidence level. It is demonstrated that the
UV treatment of an ITO plate results in a significant decrease in
the magnitude of the Casimir force (from 21\% to 35\% depending on
separation). However, ellipsometry measurements
of the imaginary parts of dielectric permittivities of the untreated
and UV-treated samples did not reveal any significant differences.
The experimental data are compared with computations in the
framework of the Lifshitz theory. It is found that the data for the
untreated sample are in a very good agreement with theoretical
results taking into account the free charge carriers in an  ITO
film. For the UV-treated sample the data exclude the theoretical
results obtained with account of free charge carriers. These data
are in a very good agreement with computations disregarding the
contribution of free carriers in the dielectric permittivity.
According to the hypothetical explanation provided, this is caused
by the phase transition of the ITO film from metallic to dielectric
state caused by the UV-treatment. Possible applications of the discovered
phenomenon in nanotechnology are discussed.
\end{abstract}

\maketitle

\section{Introduction}

Widespread interest in the van der Waals and Casimir forces
(see recent monographs\cite{1,2,3,4,5,6} and
reviews\cite{7,8,9,10,11,12}) is from the key role they play
in many physical phenomena ranging from condensed
matter physics to gravitation and cosmology.
It is common knowledge that the van der Waals force is of
quantum nature and originates from fluctuations of the
electromagnetic field. Casimir\cite{13} was the first to
generalize the van der Waals force bewtween two macrobodies
to separations where the effects of relativistic retardation
become important. The corresponding generalization of the
van der Waals interaction between an atom and a cavity wall
was performed by Casimir and Polder.\cite{14}
Both the van der Waals and Casimir forces are known under
the generic name {\it dispersion forces}.
In fact, the attractive dispersion forces between two
macrobodies, between a polarizable particle and a macrobody
and between two particles become dominant when separation
distances shrink below a micrometer. That is why these forces
are of great importance in nanotechnology where they can
play the useful role of a driving force which actuates a
microelectromechanical device.\cite{15,16}
Conversely, dispersion forces may be harmful leading
to a stable state of stiction, i.e., adhesion of free parts
of a microdevice to neighboring substrates or
electrodes.\cite{17,18}

A large body of experimental and theoretical research is
devoted to the problem of how to control the magnitude
and sign of the Casimir force. In regard to the Casimir
force with an opposite sign (the so-called {\it Casimir
repulsion}), it was possible to qualitatively demonstrate
it\cite{12} only in the case of two material bodies
separated with a liquid layer, as predicted by the
Lifshitz theory.\cite{5,6,19}
Experiments on modifying the magnitude of the attractive
Casimir force are numerous and varied. They are based
on the idea that modification of the optical properties
of the test bodies should lead to changes in the
force in accordance with the Lifshitz theory.
The Casimir force takes the largest magnitude when both
test bodies are made of good metal, e.g., of Au, which
is characterized by high reflectivity over a wide
frequency region. It was demonstrated\cite{20,21,22}
that the Casimir force between an Au sphere and a Si
plate is smaller by 25\%--40\% than in the case of two
Au bodies. This was explained by the fact that the
dielectric permittivity of Si along the imaginary frequency
axis is much smaller than that of Au. For an indium tin oxide
(ITO, In${}_2$O${}_3$:Sn) film interacting with an Au sphere
the gradient of the Casimir force was measured\cite{23,24}
to be roughly 40\%--50\% smaller than between an Au sphere
and an Au plate. In one more experiment\cite{25} it was
shown that for an Au sphere interacting with a plate made
of AgInSbTe the gradient of the Casimir force decreases
in magnitude by approximately 20\% when the material of
the plate in the crystalline phase is replaced with an
amorphous one. For a semimetallic plate, the gradient of
the Casimir force was reported to be 25\%--35\% smaller
than for an Au plate.\cite{26}

Keeping in mind the applications to micromachines, it is
important to control both decreases and increases
in the force magnitude. For this purpose, the difference
in the Casimir force between a Si plate and an Au sphere
was measured\cite{27,28} in the presence and in the
absence of 514\,nm Ar laser light on the plate.
The respective increase in magnitude of the Casimir
force by a few percent was observed when the plate is
illuminated with laser pulses. This was explained by an increase
in the charge carrier density up to five orders of magnitude
under the influence of light
 and respective changes in the
dielectric permittivity of the plate.

Investigation of the Casimir force between different materials not
only led to important experimental results with potential
applications in nanotechnology, but also raised unexpected
theoretical problems touching on the foundations of quantum
statistical physics. Thus, for two metal test bodies (an
Au-coated sphere of 150\,$\mu$m radius above an Au-coated plate) the
measured gradient of the Casimir force at laboratory temperature
was found to be in agreement with the Lifshitz theory of
dispersion forces only if the relaxation properties of free
charge carriers (electrons) are disregarded.\cite{29,30}
If these relaxation properties were taken into account by means of
the Drude model, the Lifshitz theory was found to be excluded by
the experimental data at almost 100\% confidence level.
Recently it was claimed\cite{31} that an experiment using an
Au-coated spherical lens of 15.6\,cm curvature radius above an
Au-coated plate is in agreement with theory taking into account
the relaxation properties of free electrons. The results of this
experiment are in contradiction with the above mentioned
measurement using a small sphere\cite{29,30} and another
experiment using large spherical lens.\cite{32}
(See Refs.\cite{33,34} for detailed critical discussion.)

For an Au sphere interacting with a Si plate illuminated with
laser pulses the experimental data for the difference  in the
Casimir forces in the presence and in the absence of light
agree with the Lifshitz theory only if the charge carriers of
dielectric Si in the absence of light are disregarded.\cite{27,28}
When the charge carriers are taken into account, the Lifshitz
theory is excluded by the data at a 95\% confidence level.
Similar results were obtained from the measurement\cite{35}
of the thermal Casimir-Polder force between ${}^{87}$Rb atoms
belonging to the Bose-Einstein condensate and a SiO${}_2$ plate:
theory is in agreement with the data when dc conductivity
of the dielectric plate is disregarded,\cite{35}
but the same theory is excluded by the data at a 70\%
confidence level when dc conductivity is taken into
account.\cite{36}

In this paper, we present complete experimental
and theoretical results for
the Casimir force between an Au-coated sphere and ITO films
deposited on a quartz substrate. Measurements were performed
using a modified multimode atomic force microscope (AFM) in
high vacuum. The main difference of this experiment in
comparison with Refs.\cite{23,24} is that in two sets of
measurements the ITO sample was used as is, but
another two sets of
measurements were done after the sample was subjected to
 UV treatment. Unexpectedly, it was observed that the UV
treatment results in a significant decrease in the
magnitude of the Casimir force (from 21\% to 35\% at
different separations). This decrease is not associated with
respective modifications of the optical properties of plates
under the influence of UV treatment, as was confirmed by
means of ellipsometry measurements (preliminary results of
this work based on only one data set were published in
Ref.\cite{38}).

The experimental results are compared with calculations
using the Lifshitz theory and different models of the
dielectric properties of the test bodies. Note that ITO
at room temperature is a good conductor at quasistatic
frequencies, but is transparent to visible and near
infrared light. Keeping this in mind, it was
suggested\cite{39} to use this material in investigations
of the Casimir force. Computations are done for a
four-layer system (ITO on quartz interacting with Au
through a vacuum gap). The experimental results for an
untreated ITO sample are found to be in agreement with
the Lifshitz theory if charge carriers are taken into
account. For a UV-treated sample, the Lifshitz theory
taking into account the charge carriers is excluded
by the experimental data at a 95\% confidence level.
These data are found consistent with computations
disregarding the charge carriers in the ITO sample.
Based on this, the hypothesis is proposed that the
UV treatment resulted in the transition of the ITO
film to a dielectric state without noticeable
change of its dielectric permittivity at the
laboratory temperature.

The paper is organized as follows. In Sec.~II we
describe the experimental setup used and the procedures
of sample preparation and characterization.
Section~III contains details of the electrostatic
calibrations. This includes determination of the residual
potential difference, the deflection coefficient,
the separation on contact,
and the calibration constant.
In Sec.~IV the experimental results for the Casimir
force are presented. We consider both the individual
and mean measured forces and calculate the random,
systematic and total experimental errors for an
untreated and UV-treated samples at a 95\% confidence
level.
Section~V is devoted to the comparison between
experiment and theory. Here, special attention
is paid to the complex refractive indices and
dielectric permittivities along the imaginary
frequency axis used in the computations.
All computational results are obtained in the
framework of different approaches to the problem
of free charge carriers in the Lifshitz theory.
In Sec.~IV the reader will find our conclusions
and discussion.

\section{Experimental setup and sample characterization}

All measurements of the total force (electrostatic plus
Casimir) between an Au-coated polystyrene sphere and
an ITO film on a quartz plate were performed using
a modified multimode AFM in a high vacuum.
In this section we describe the most important details
of the experimental setup and procedures used for a
sample preparation and characterization.

\subsection{Schematic of the experimental setup}

The commercial AFM (``Veeco") used in our measurements
was modified to be free of volatile organics. It was placed in a
high vacuum chamber (see Fig.~1). Only oil-free mechanical and
turbo pumps shown in Fig.~1 were used to obtain the vacuum.
As a result, the experiments were done at a pressure of
$10^{-6}\,$Torr. To ensure a low vibration noise environment,
we used an optical table and a sand damper box to prevent
coupling of the low frequency noise from the mechanical and
turbo pumps (see Fig.~1).

After the first use\cite{40} of the AFM to measure the Casimir
force, it has been employed for this purpose in many
experiments.\cite{20,21,22,23,24,25,26,27,28,41,42,43,44,45}
The AFM system consists of a head, piezoelectric actuator,
an AFM controller and computer. The head includes a diode
laser which emits a collimated beam with a waist of tens
of micrometers at the focus,
an Au-coated cantilever with attached sphere
that bends in response to the sphere-plate force, and
photodetectors which measure the cantilever deflection
through a differential measurement of the laser beam intensity.
The plate is mounted on the top of the piezoelectric actuator
which allows movement of the plate towards the sphere for
a distance of $2\,\mu$m. To change the sphere-plate distance
and avoid piezo drift and creep, a continuous 0.05\,Hz
triangular voltage signal was applied to this actuator.
The application of the voltages to both the piezoelectric
actuator and the laser, and the electronic processing and
digitizing of the light collected by the photodetectors
is done by the AFM controller (see Fig.~1).
The computer is used for data acquisition.
The cantilever deflection was recorded for about every
0.2\,nm movement of the piezoelectric actuator.

More specifically, the scheme of the experiment is as follows.
A total force $F_{\rm tot}$ acting between a sphere and
a plate causes an elastic deflection $z$ of the cantilever
in accordance with Hooke's law
\begin{equation}
F_{\rm tot}=kz,
\label{eq1}
\end{equation}
\noindent
where $k$ is the spring constant. When the separation distance
between the sphere and the plate is changed with the movement
of the piezoelectric element, the cantilever deflection will
correspond to the different forces it experiences at
different separations. This deflection causes the deviation
of the laser beam reflected off the cantilever tip which is
measured with photodiodes A and B (see Fig.~1).
The respective deflection signal $S_{\rm def}$ at various
separation distances leads to a force-distance curve.
Note that the signal $S_{\rm def}$ recorded by the photodetector
is not in force units but has to be calibrated according to
Eq.~(\ref{eq1}) and
\begin{equation}
z=mS_{\rm def},
\label{eq2}
\end{equation}
\noindent
where $m$ is the cantilever deflection in nm per unit photodetector
signal (sometimes called the optical lever sensitivity).
Here $S_{\rm def}$ is measured in volts and $m$ in nm/V.

To stabilize the laser used for the detection of deflection of
the AFM cantilever, we employed a liquid nitrogen cooling system,
which maintained the temperature of the AFM at 2${}^{\circ}$C.
This is the temperature at which all measurements were performed
in this experiment.
We attached a copper braid to the surface of the AFM laser soarce.
The other side of the braid was attached to a liquid nitrogen
reservoir, which was also located inside the vacuum chamber.
During the experiments the reservoir could be refilled through
a liquid feed-through (see Fig.~1). The cooling system helped us
to improve the spot size and to reduce the laser noise and drift.
It also served as an additional cryo pump to obtain the high
vacuum.

\subsection{Sample preparation and characterization}

The test bodies in this experiment consisted of an Au-coated sphere
and quartz plate coated with an ITO film. Measurements of the
Casimir force were performed for a sphere interacting with an
untreated or, alternatively, an UV-treated ITO sample.
First, we briefly consider the preparation of the sphere which was
done similar to previous
experiments.\cite{20,21,22,27,28,40,41,44,45}

We used a polystyrene sphere which was glued with silver epoxy
($20\times 20\,\mu\mbox{m}^2$ spot) to the tip of a triangular
silicon nitride cantilever with a nominal spring constant of
order 0.01\,N/m. The cantilever-sphere system was then coated
with a 10\,nm Cr layer followed by 20\,nm Al layer and finally
with a $105\pm 1\,$nm Au layer. This was done in an oil free
thermal evaporator at a $10^{-7}\,$Torr vacuum.
To make sure that the Au surface is sufficiently smooth,
the coatings were performed at a very low deposition rate of
3.75\,\AA/min.
The radius of an Au-coated sphere was determined using a
scanning electron microscope to be $R=(101.2\pm 0.5)\,\mu$m.
This was done after the end of force measurements.

The ITO film used in our experiment was prepared by RF sputtering
(Thinfilm Inc.) on a 1\,cm square single crystal quartz plate of
1\,mm thickness. The film thickness and nominal resistivity were
measured to be $d=(74.6\pm 0.2)\,$nm and $42\,\Omega$/sq, respectively.
The surface of the ITO film was cleaned using the following
procedure. First, the ITO sample was immersed in acetone and
cleaned in an ultrasonic bath for 15\,min. Then it was rinsed
3 times in DI water. This ultrasonic cleaning procedure and water
rinsing was repeated next with methanol followed by ethanol.
After completing of the ultrasonic cleaning, the sample was dried
in a flow of pure nitrogen gas. Next, electrical
contacts to copper wires were made by soldering with an indium
wire. The ITO sample was now ready for force
measurements in the high vacuum chamber which are done as described
below.

After the force measurements were completed, the ITO sample was
UV treated. For this purpose it was placed in a special air
chamber containing a UV lamp. A pen-ray Mercury lamp with
a length of $9.0^{\prime\prime}$ and outside diameter of
$0.375^{\prime\prime}$ was used as
the UV source. This lamp emits a spectrum with the primary peak
at the wavelength 254\,nm (5.4\,mW/cm${}^2$ at 1.9\,cm
distance) and a secondary peak at 365\,nm (0.2\,mW/cm${}^2$ at 1.9\,cm
distance). During the UV treatment the sample was placed at 1\,cm
from the lamp for 12 hours. After finish of the UV treatment,
the sample was cleaned as described above, and then the force
measurements were again performed.

An important part of surface characterization is the measurement of
roughness profiles on both an ITO sample and an Au-coated sphere.
The roughness of the untreated and UV-treated
ITO samples and the Au-coated sphere was
investigated using the AFM.
For the ITO plate before and after the UV treatment the roughness was
found to be the same.
A typical three-dimensional scan of an
ITO plate is shown in Fig.~2(a). As can be seen in this figure,
the roughness is represented by stochastically distributed
distortions. For comparison purposes, the two-dimensional AFM
scan of the surface of an ITO sample is shown in Fig.~2(b).
Here, the lighter tone corresponds to the larger height above the
minimum roughness level. The analysis of the data of AFM scans allows
to determine the fraction of plate area $v_i^{\rm (ITO)}$ with
heights $h_i^{\rm (ITO)}$ where $i=1,\,2,\,\ldots,\,N_1$ and $N_1$
is some chosen number. These heights are measured from the
absolute minimum level on the test body $h_1^{\rm (ITO)}=0$.
The resulting distribution function for an ITO plate is shown
in Fig.~3(a). The zero roughness level on an ITO sample,
relative to which the mean value of the roughness is equal to
zero,\cite{6,10} takes the value $H_0^{\rm (ITO)}=9.54\,$nm
with $N_1=18$. The respective variance describing the stochastic
roughness on an ITO sample\cite{6,10} is given by
$\delta_{\rm ITO}=2.28\,$nm. The distribution function for the
roughness on an Au-coated sphere
[$v_i^{\rm (Au)}$ as a function of $h_i^{\rm (Au)}$]
is presented in Fig.~3(b).
Here, the corresponding zero roughness level and variance are
given by $H_0^{\rm (Au)}=11.51\,$nm and $\delta_{\rm Au}=3.17\,$nm,
respectively, with $N_2=25$.

The presence of roughness on the surface determines the minimum
separation distance that can be achieved when the test bodies
are approaching. In fact the minimum separation $z_0$ (the
so-called separation on contact) is the separation between the zero
levels of the roughness on contact of the two surfaces.
The actual absolute separation between the zero roughness levels
on the bottom of an Au sphere and the ITO plate with account of
Eq.~(\ref{eq2}) is given by
\begin{equation}
a=z_0+z_{\rm piezo}+mS_{\rm def},
\label{eq3}
\end{equation}
\noindent
where $z_{\rm piezo}$ is the distance moved by the plate owing to
the voltage applied to the piezoelectric actuator.
Figure~4 illustrates the meaning of the average separation on
contact $z_0$ and other parameters entering Eq.~(\ref{eq3}).
In Sec.~VB the roughness profiles will be used to calculate
the theoretical values of the Casimir force.

\section{Electrostatic calibrations}

Electrostatic calibrations in the measurements of the Casimir
force need extreme care. They allow determination with
sufficient precision of values of such vital parameters as
the residual potential difference $V_0$, the separation on
contact $z_0$, the spring constant $k$, and the cantilever
deflection coefficient $m$. During calibration process
it is necessary to make sure that all relevant electric forces
acting in the experimental configuration are included in the
theoretical model used, and all possible background forces
are negligibly small (see, for instance, a
discussion\cite{46,47,48,49} on the role of patch potentials which
may exist due to the grain structure of metal coatings, surface
contaminants etc.).

To make electrostatic measurements, the ITO plate was connected to
a voltage supply (33120A,``Agilent Inc.'') operating with $1\,\mu$V
resolution, while the sphere remained grounded. A 1\,k$\Omega$
resistor was connected in series with the voltage supply to
prevent surge currents and protect the sample surface during
sphere-plate contact.
The cantilever-sphere system was mounted on the AFM head which
was connected to the ground. To reduce the electrical noise,
care was taken to make Ohmic contacts and eliminate all Schottky
barriers to the ITO plate and Au sphere. To minimize electrical
ground loops, all the electrical ground connections were
unified to the AFM ground. Ten different voltages $V_i$
($i=1,\,2,\,\ldots,\,10$) were applied to the plate in each
round of measurements. For an untreated plate these voltages were
in the range from --260 to --110\,mV in the first set of
measurements and from --265 to --115\,mV in the second set of
measurements. For a UV-treated plate the applied voltages were from
--25 to 150\,mV and from --5 to 140\,mV in the first and second
set of measurements, respectively.

The total force between the sphere and the plate is given by the
sum of electric $F_{\rm el}$ and Casimir $F$ forces.
In accordance with Eqs.~(\ref{eq1}) and (\ref{eq2}) it can be
represented in the form
\begin{equation}
F_{\rm tot}(a,V_i)=F_{\rm el}(a,V_i)+F(a)=
\tilde{k}S_{\rm def}(a,V_i),
\label{eq4}
\end{equation}
\noindent
where $\tilde{k}=km$ is the calibration constant.
This force was measured as a function of separation.
As described in Sec.~IIA, to change the separation a continuous
triangular voltage was applied to the AFM piezoelectric actuator.
Note that this piezoelectric actuator was calibrated
interferometrically.\cite{50,51}
Starting at the maximum separation of $2\,\mu$m, the ITO plate was
moved towards the Au sphere and the corresponding cantilever
deflection was recorded at every 0.2\,nm until the plate
contacted the sphere.

After the contact of the sphere and the plate, the cantilever-sphere
system vibrated with a large amplitude. To allow time for this
vibration to damp out, a 5\,s delay was introduced after every cycle
of data acquisation. To reduce random error, the total force at
each of ten voltages applied to the untreated plate was measured ten
times as a function of separation. The same was repeated for the
UV-treated plate.

The electric force between a sphere and a plate made of conductors is
given by\cite{6,52}
\begin{equation}
F_{\rm el}(a,V_i)=X(a)(V_i-V_0)^2,
\label{eq5}
\end{equation}
\noindent
where
\begin{eqnarray}
&&
X(a)=2\pi\epsilon_0\sum_{n=1}^{\infty}
\frac{\coth\alpha-n\coth{n\alpha}}{\sinh{n\alpha}},
\nonumber \\
&&
\cosh\alpha=1+\frac{a}{R},
\label{eq6}
\end{eqnarray}
\noindent
$V_0$ is the residual potential difference which can be present
due to different work functions of the sphere and plate materials,
and $\epsilon_0$ is the permittivity of the vacuum.
In the wide range of separations the function $X(a)$ can be
presented in the polynomial form\cite{21}
\begin{equation}
X(a)=-2\pi\epsilon_0\sum_{i=-1}^{6}c_i
\left(\frac{a}{R}\right)^i
\label{eq7}
\end{equation}
\noindent
with an error of about 0.01\%. The coefficients $c_i$ in
Eq.~(\ref{eq7}) are given by
\begin{eqnarray}
&&
c_{-1}=0.5,\ \  c_0=-1.18260,\ \  c_1=22.2375,
\nonumber \\
&&
c_2=-571.366,\ \
c_3=9592.45, \ \  c_4=-90200.5,
\nonumber \\
&&
c_5=383084, \ \  c_6=-300357.
\label{eq8}
\end{eqnarray}

As can be seen in Eqs.~(\ref{eq4}) and (\ref{eq5}), the total
force $F_{\rm tot}$ is characterized by the parabolic dependence
of an applied voltage $V_i$. The same is true for the total
deflection signal
\begin{equation}
S_{\rm def}(a,V_i)=S(a)+\frac{X(a)}{\tilde{k}}
(V_i-V_0)^2,
\label{eq9}
\end{equation}
\noindent
where $S(a)=F(a)/\tilde{k}$ is the deflection due to the
Casimir force.

As an example, in Fig.~5 we present as squares the measured deflection
signal $S_{\rm def}$ plotted as a function of the applied voltage for
(a) the untreated and (b)  UV-treated sample at a fixed separation
$a=75\,$nm between the sphere and the plate.
Then a $\chi^2$ fitting procedure was used to draw parabolas
[the solid lines in Fig.~5(a,b)] and to determine their vertices
and the coefficients $\beta(a)=X(a)/\tilde{k}$.
 This procedure was repeated at each
separation distance with a step of 1\,nm
(though data were acquired about every 0.2\,nm of $z_{\rm piezo}$,
only interpolated values at 1\,nm step were analyzed).
The vertex of each parabola corresponds to $V_0$ at each respective
separation. When $V_i=V_0$, the electrostatic force is equal to zero.
The fitting procedure was also repeated at every separation $a$.
{}From the parabolas shown in Fig.~5(a,b) it was obtained
$V_0=-(195.9\pm 0.5)\,$mV and $V_0=(64.7\pm 0.4)\,$mV, respectively.
The values of $V_0$ at separations from 60 to 300\,nm are shown in
Fig.~6(a) for an untreated and in Fig.~7(a) for a UV-treated sample
in the first measurement set.

As can be seen in Figs.~6(a) and 7(a),  there are
anomalous dependences of $V_0$ on $a$,
i.e., $V_0$ is not constant as expected if the electric force
is described by the exact Eqs.~(\ref{eq5}) and (\ref{eq6}).
Such dependences were found in several experiments measuring the
Casimir force and widely discussed in the literature.\cite{46,47,48,49,55}
They are often interpreted as a manifectation of an additional electric
force due to the presence of electrostatic surface impurities and space
charge effects on the sphere or plate surfaces.
In our case, however, these seeming anomalies do not indicate the
presence of some extra electric force other than that given by
Eqs.~(\ref{eq5}) and (\ref{eq6}). The point is that the preceding
analysis did not take into consideration the finiteness of the data
acquisition rate and, more importantly, the mechanical drift of the
sphere-plate separation. As a result, systematic
deviations occurred in
the residual potential difference $V_0$ in Figs.~6(a) and 7(a).
These systematic deviations have been investigated,\cite{53,54} and the
respective corrections have been introduced in the data for $V_0$
as a function of $a$.
The corrected results for $V_0$ as a function of separation for an
untreated sample are shown in Fig.~6(b) and for a UV-treated sample
in Fig.~7(b). Specifically, at $a=75\,$nm [see Fig.~5(a,b)] the
corrected values of $V_0$ for an untreated and UV-treated samples are
$V_0=-(195.6\pm 0.5)\,$mV and $V_0=(64.4\pm 0.4)\,$mV, respectively.
As is seen in Figs.~6(b) and 7(b), after the proper corrections
for the mechanical drift of separation distances and the finitness
of the data acquisition rate are introduced, the residual potential
difference remains constant in the limits of random errors.
This excludes the presence of any perceptible electric force
due to dust and contaminants in
addition to the one given by Eqs.~(\ref{eq5}) and (\ref{eq6}).
{}From Figs.~6(b) and 7(b) the following mean values for the residual
potential difference were found:
$V_0=-(196.8\pm 1.5)\,$mV for the untreated sample and
$V_0=(65\pm 2)\,$mV for the UV-treated sample.

We now discuss the deflection coefficient $m$ which enters
Eq.~(\ref{eq3}) and, thus, is needed to determine both the absolute
and relative sphere-plate separations.   The
deflection signals obtained by the application of the different
voltages to the plate can be used to determine $m$.  Here larger
$(V-V_0)^2$ will lead to correspondingly larger deflections and thus
sphere-plate contact at smaller $z_{\rm piezo}$.
This rate of change of
contact point with $S_{\rm def}$ gives the value of $m$.
However, as before, care must be
taken to make a precise determination of the point of
sphere-plate contact and correct the contact point for mechanical
drift of the sphere-plate separation.
Both corrections are already discussed in detail.\cite{53,54}
Briefly, the first of them is necessary, as even at the maximum
acquisition rate, data points are widely separated near the point of
contact due to the large force gradient at short separations. Thus an
interpolation procedure has to be used to determine the exact contact
point.  With respect to the second correction,
the contact points with two different applied $V$ but same
$(V-V_0)^2$ must be the same as the corresponding cantilever
deflections are equal.
Because of this any observed  change in the contact point
for these two applied voltages is due to sphere-plate drift. The
drift rate is the time rate of change in the contact points
for these voltages. Both
corrections are needed to obtain the precise relative sphere-plate
separation. As we discussed above,
neglect of the drift correction leads to anomalous
distance dependence behavior for the residual potential $V_0$.
The corrected values of
$m=(104.4\pm 0.5)\,$nm/V
for the untreated sample and
$m=(103.5\pm 0.6)\,$nm/V
for the UV-treated sample were determined
for the first measurement set.

In accordance with Eqs.~(\ref{eq3}) and (\ref{eq7}), the parabola
coefficient $\beta(a)$ depends on both the cantilever calibration
constant $\tilde{k}$ and the average separation on contact $z_0$.
Thus, the obtained values of $\beta(a)$ at different separations
can be used to determine both $z_0$ and $\tilde{k}$ using the least
$\chi^2$-fitting to Eq.~(\ref{eq7}) as described previously.\cite{53,54}
For this purpose the coefficient $\beta(a)$ was first fitted from a
starting point of 60\,nm to an end point $a_{\rm end}=1000\,$nm, and the
values of $z_0$ and $\tilde{k}$ were determined. Then $a_{\rm end}$
was decreased to 900\,nm and the fitting procedure repeated leading
to the corresponding values of $z_0$ and $\tilde{k}$.
The repetition of this procedure (in smaller steps below
$a_{\rm end}=400\,$nm) results in Figs.~8(a) and 9(a) demonstrating
the dependence of $z_0$ on $a_{\rm end}$ for an untreated and UV-treated
samples, respectively, in the first set of
our measurements. Similar results
were obtained for $\tilde{k}$ as a function of $a_{\rm end}$.

As can be seen in Figs.~8(a) and 9(a), there is an anomalous
dependence of $z_0$ on $a_{\rm end}$ caused, as above, by mechanical
drift in the sphere-plate separation.
After the respective corrections were introduced,\cite{53,54}
the obtained dependence
of the separation on contact $z_0$ on $a_{\rm end}$ is shown in
Fig.~8(b) for the untreated sample and in Fig.~9(b) for the UV-treated
sample. It can be seen that in both cases the corrected values
$z_0$ remain constant within the limits of random errors
independently of the value of $a_{\rm end}$ chosen.
The obtained mean values of $z_0$ are
$z_0=(29.5\pm 0.4)\,$nm
for the untreated sample and
$z_0=(29.0\pm 0.6)\,$nm
for the UV-treated sample.
In Fig.~10(a,b) the corrected values of the calibration constant
$\tilde{k}$ as a function of $a_{\rm end}$ obtained in the first
measurement set are shown for an untreated and UV-treated samples,
respectively. The respective mean values are as follows:
$\tilde{k}=(1.45\pm 0.02)\,$nN/V and
$\tilde{k}=(1.43\pm 0.02)\,$nN/V.
As is seen in Fig.~10(a,b), the individual values of $\tilde{k}$
determined with different $a_{\rm end}$ are constant in the limits
of random errors.
This concludes the electrostatic calibration of our measurement
system. It is pertinent to note that the fitting used above was
made to only the well understood electric force in the sphere-plate
configuration. In so doing it was confirmed that there are no other
perceptible electric forces due to surface patches, contaminants,
surface defects, such as pits and bubbles, etc.

The same calibration procedure, as described above, was repeated
when performing the second set of our measurements.
For the untreated sample, the following values of the parameters
were found:
$V_0=-(196.8\pm 1.5)\,$mV,
$z_0=(29.6\pm 0.5)\,$nm,
$\tilde{k}=(1.51\pm 0.02)\,$nN/V, and
$m=(104.4\pm 0.5)\,$nm/V.
For the UV-treated sample it was obtained:
$V_0=(64.8\pm 2)\,$mV,
$z_0=(29.0\pm 0.6)\,$nm,
$\tilde{k}=(1.51\pm 0.02)\,$nN/V, and
$m=(104.2\pm 0.6)\,$nm/V.
The parameters presented in this section were used to convert
the cantilever deflection signals into the values of the total
force and to find the values of absolute separations.

Note that for all values of the above parameters  the errors are
indicated at a 95\% confidence level.

\section{Measurement results for the Casimir force and
their errors}

According to Eq.~(\ref{eq4}) the experimental results for the
Casimir force between an Au sphere and an ITO plate are given by
\begin{equation}
F(a)=\tilde{k}S_{\rm def}(a,V_i)-F_{\rm el}(a,V_i),
\label{eq10}
\end{equation}
\noindent
where the electric force $F_{\rm el}$ is expressed by
Eqs.~(\ref{eq5})--(\ref{eq7}).
At each separation $a$ the measurement of $S_{\rm def}$ with
ten applied voltages was performed. This was repeated 10 times.
Altogether, 100 values of the Casimir force at each separation
were obtained from Eq.~(\ref{eq10}) in each measurement set for
both untreated and UV-treated samples. Here, we present the main
features of these data and determine the experimental errors.

\subsection{Mean measured Casimir forces}

In Fig.~11 the mean measured Casimir forces obtained from one
hundred individual values are shown as functions of separation
with solid lines for (a) an untreated sample and (b) a
UV-treated sample over the separation region from 60 to 300\,nm
in the first measurement set. In the insets, the same solid lines
are reproduced over a more narrow separation region from 60 to 100\,nm.
As an illustration, Fig.~11 shows by dots all 100 individual values of
the Casimir force plotted at separation distances with a step of 5\,nm
(in the insets with a step of 1\,nm). Figure~11(a) indicates a
40\%--50\% decrease in the force magnitude in comparison with the
case of two Au bodies in agreement with previous work\cite{23,24}
where a similar result was obtained for the Casimir pressure.
For example at $a=80\,$nm the measured Casimir force is --144\,pN
in contrast to --269\,pN for Au test bodies.
As can be seen in Fig.~11(a,b), the magnitudes of the Casimir force
from a UV-treated plate are 21\% to 35\% smaller than from
an untreated plate.

Figure~12 characterizes the statistical properties of the experimental
data for an untreated (right) and UV-treated (left) samples by
presenting the histograms for the measured Casimir force at
separations (a) $a=60\,$nm, (b) $a=80\,$nm, and (c) $a=100\,$nm
in the first measurement set. The histograms are described by
Gaussian distributions (dashed lines) with the standard deviations
equal to
(a) $\sigma_{G}=4.6\,$pN (right), $\sigma_{G}=5.0\,$pN (left),
(b) $\sigma_{G}=5.4\,$pN (right), $\sigma_{G}=5.4\,$pN (left),
and
(c) $\sigma_{G}=4.9\,$pN (right), $\sigma_{G}=4.7\,$pN (left).
The values of the respective mean measured Casimir forces can be found
in columns 2 and 5 of Table~I.
{}From Fig.~12 it is observed that the Gaussian distributions related
to the untreated and UV-treated samples do not overlap lending great
confidence to the effect of a decrease of the magnitude of the
Casimir force under the influence of UV-treatment.

In Table~I we present the mean magnitudes of the measured Casimir
force at different separations (first column) ranging from 60 to
300\,nm with the respective total experimental errors determined below
in Sec.~IVB. Columns 2 and 3 contain the force magnitudes obtained
in the first and second measurement sets for the untreated sample,
respectively. In columns 5 and 6 the respective results obtained for
the UV-treated sample in the first and second measurement sets are
presented. As can be seen in Table~I, the measurement data obtained
in the first and second measurement sets are in a very good
agreement. All differences between them are much less than the total
experimental errors presented in columns 4 and 7
(see Sec.~IVB).

\subsection{Random, systematic and total experimental errors}

Here we present the main results of the error analysis.
The variance of the mean Casimir force calculated from 100 measurement
results over the separation interval from 60 to 300\,nm is shown as
dots in Figs.~13(a) and 13(b) for the untreated and UV-treated samples,
respectively. The respective mean values are separation independent:
$\sigma=0.55\,$pN and $\sigma=0.5\,$pN.
They are equal to the random errors in the measured Casimir force
determined at a 67\% confidence level. To determine the random error
at a $\beta=95$\% confidence level, one should multiply $\sigma$ by
the student coefficient $t_{1.95/2}(99)=2$. Thus, the random errors
at a 95\% confidence level are equal to
$\Delta^{\!r}F=1.1\,$pN and $\Delta^{\!r}F=1.0\,$pN, respectively.

According to Eq.~(\ref{eq10}), the systematic error in the measured
Casimir forces is a combination of the systematic errors in the total
measured force and subtracted electric force. The systematic error in
the total measured force, $\Delta^{\! s}F_{\rm tot}$, is determined
by the instrumental noise including the background noise level, and
the errors in calibration. In Figs.~14(a) and 14(b) the error
$\Delta^{\! s}F_{\rm tot}$ determined at a 95\% confidence level is
shown by the long-dashed lines as a function of separation for the
untreated and UV-treated samples, respectively. The error in
calculation of the electric force, which plays the role of a systematic
error with respect to the Casimir force obtained from Eq.~(\ref{eq10}),
is mostly determined by the errors in the measurement of separations.
The latter are largely contributed by the errors in $z_0$ presented
in Sec.~III.
As a result, for the first measurement set, the errors in absolute
separations determined at a 95\% confidence level are equal to
$\Delta a=0.4\,$nm and $\Delta a=0.6\,$nm for the untreated and
UV-treated samples, respectively.
Note that due to Eq.~(\ref{eq5}) the error in $F_{\rm el}$
is different at different applied voltages $V_i$. As an illustration,
Fig.~14(a,b) shows with short-dashed lines the mean
$\Delta^{\! s}F_{\rm el}$ averaged over 10 applied voltages for the
untreated and UV-treated samples, respectively. The respective solid
lines in Figs.~14(a) and 14(b) demonstrate the total systematic
errors $\Delta^{\! s}F$ in the Casimir force as a function of separation.
They were obtained by adding in quadrature the systematic errors of the
total and electric forces.
As is seen in Fig.~14(a,b), at moderate and large separations the
major contribution to the systematic error in the Casimir force is
given by the systematic error in the total force. Only at short
separations the error in the electric force contributes significantly
to the systematic error in the Casimir force.

To obtain the total experimental error one should combine the random
and systematic errors. In Fig.~15(a,b) the random errors are shown
with the dashed lines for the untreated and UV-treated samples, respectively.
The lower solid lines in the same figure represent the systematic errors
which are dominant over the random ones in this experiment, especially
at short separations. Keeping in mind that both the random and
systematic errors considered above are characterized by the normal
distribution, they should be added in quadrature.
The resulting absolute total experimental errors
$\Delta^{\!\rm tot}F$ determined at a 95\% confidence
level are shown by the upper solid lines for the untreated [Fig.~15(a)]
and UV-treated [Fig.~15(b)] samples. The values of the absolute total
experimental errors at different separations are presented in column~4
of Table~I (for an untreated sample) and in column~7 for a
UV-treated sample. The relative total experimental error in the measured
Casimir force at $a=60\,$nm is equal to 0.82\% and 1.2\% for the
untreated and UV-treated samples, respectively. With the increase of
separation to $a=100\,$nm the respective errors increase to 2.3\%
and 3.6\% and further increase to 13.4\% and 24.2\% when separation
increases to $a=200\,$nm. At $a=300\,$nm the
relative total experimental
errors in the measured Casimir force for the untreated and UV-treated
samples achieve 37.5\% and 50\%, respectively.
For the sake of definiteness, these numerical values are given for
the first measurement set. However, in both sets the total
experimental errors are the same.

\section{Comparison between experiment and theory}

We next discuss the comparison between the experimental data
for the Casimir force and the theoretical predictions of the
Lifshitz theory. We compute theoretical results using different
approaches to the thermal Casimir force proposed in the
literature.\cite{6,8,10}
In the framework of the proximity force approximation which is
clearly applicable\cite{57} for the parameters of this experiment
the Lifshitz formula for the Casimir force between an Au-coated
sphere and an ITO film deposited on a quartz plate takes the form
\begin{eqnarray}
F(a,T)&=&
k_BTR\sum_{l=0}^{\infty}{\vphantom{\sum}}^\prime\int_0^{\infty}
k_{\bot}dk_{\bot}
\nonumber \\
&\times&
\left\{\ln\left[1-
r_{\rm TM}^{(1)}(i\xi_l,k_{\bot})
r_{\rm TM}^{(2)}(i\xi_l,k_{\bot})\,e^{-2aq_l}\right]\right.
\nonumber \\
&&+\left.
\ln\left[1-
r_{\rm TE}^{(1)}(i\xi_l,k_{\bot})
r_{\rm TE}^{(2)}(i\xi_l,k_{\bot})\,e^{-2aq_l}\right]\right\}.
\label{eq11}
\end{eqnarray}
\noindent
Here, $k_B$ is the Boltzmann constant, $T=275\,$K,
$\xi_l=2\pi k_BTl/\hbar$ with $l=0,\,1,\,2,\,\ldots$ are
the Matsubara frequencies, the primed sum means that the term
with $l=0$ is divided by 2, $k_{\bot}$ is the projection of the
wave vector onto the plate, and $q_l^2=k_{\bot}^2+\xi_l^2/c^2$.
The reflection coefficients on an Au body modeled as a semispace
for the transverse magnetic (TM) and transverse electric (TE)
polarizations of the electromagnetic field are given by
\begin{eqnarray}
&&
r_{\rm TM}^{(1)}(i\xi_l,k_{\bot})=
\frac{\varepsilon_l^{(1)}q_l-k_l^{(1)}}{\varepsilon_l^{(1)}q_l+
k_l^{(1)}},
\nonumber \\
&&
r_{\rm TE}^{(1)}(i\xi_l,k_{\bot})=
\frac{q_l-k_l^{(1)}}{q_l+k_l^{(1)}},
\label{eq12}
\end{eqnarray}
\noindent
where ${k_l^{(1)}}^2=k_{\bot}^2+\varepsilon_l^{(1)}\xi_l^2/c^2$ and
$\varepsilon_l^{(1)}\equiv\varepsilon^{(1)}(i\xi_l)$ is the dielectric
permittivity of Au along the imaginary frequency axis.

The reflection coefficients of an ITO film deposited on quartz plate
can be presented in the form\cite{6,58}
\begin{equation}
r_{\rm TM}^{(2)}(i\xi_l,k_{\bot})=
\frac{r_{\rm TM}^{(0,-1)}+
r_{\rm TM}^{(-1,-2)}e^{-2k_l^{(-1)}d}}{1+
r_{\rm TM}^{(0,-1)}r_{\rm TM}^{(-1,-2)}e^{-2k_l^{(-1)}d}}
\label{eq13}
\end{equation}
\noindent
and the same expression with the index TM replaced for TE.
Here, $r_{\rm TM,TE}^{(n,n')}$ are
the reflection coefficients on an
ITO layer of thickness $d$ ($n=0,{\ }n'=-1$) and on a thick
quartz plate modeled as a semispace ($n=-1,{\ }n'=-2$)
 \begin{eqnarray}
&&
r_{\rm TM}^{(n,n')}(i\xi_l,k_{\bot})=
\frac{\varepsilon_l^{(n')}k_l^{(n)}-
\varepsilon_l^{(n)}k_l^{(n')}}{\varepsilon_l^{(n')}k_l^{(n)}+
\varepsilon_l^{(n)}k_l^{(n')}},
\nonumber \\
&&
r_{\rm TE}^{(n,n')}(i\xi_l,k_{\bot})=
\frac{k_l^{(n)}-k_l^{(n')}}{k_l^{(n)}+k_l^{(n')}}.
\label{eq14}
\end{eqnarray}
\noindent
The notations used are the following:
$\varepsilon_l^{(0)}=1$,
$\varepsilon_l^{(-1)}=\varepsilon^{(-1)}(i\xi_l)$ and
$\varepsilon_l^{(-2)}=\varepsilon^{(-2)}(i\xi_l)$ are
the dielectric permittivities of ITO and quartz, respectively,
and ${k_l^{(n)}}^2=k_{\bot}^2+\varepsilon_l^{(n)}\xi_l^2/c^2$.
To perform computations of the Casimir force using
Eqs.~(\ref{eq11})--(\ref{eq14}) one needs the dielectric
permittivities of Au, ITO and quartz over a wide range of
imaginary frequencies.

\subsection{Complex indices of refraction and dielectric
permittivities
along imaginary frequencies}

We describe the imaginary part of the dielectric permittivity of Au,
${\rm Im}\,\varepsilon^{(1)}(\omega)$, by means of the tabulated
optical data.\cite{59} In the region $\omega<0.125\,$eV, where the
optical data are missing, the extrapolation by means of the
imaginary part of the Drude model dielectric permittivity with
the plasma frequency $\omega_p=9.0\,$eV and relaxation parameter
$\gamma=0.035\,$eV has been used. The dielectric permittivity
of Au along the imaginary frequencies was obtained by means of
the Kramers-Kronig relations.\cite{6} Using the so-called
weighted Kramers-Kronig relations it was recently shown\cite{60}
that the extrapolation by means of the Drude model with
$\omega_p$ and $\gamma$ indicated above is in excellent agreement
with the optical data measured over a wide frequency region.

For the underlying quartz plate we used the averaged dielectric
permittivity obtained\cite{61} in the Ninham-Parsegian
approximation\cite{5}
\begin{equation}
\varepsilon^{(-2)}(i\xi_l)=1+
\frac{C_{\rm IR}}{1+\frac{\xi_l^2}{\omega_{\rm IR}^2}}+
\frac{C_{\rm UV}}{1+\frac{\xi_l^2}{\omega_{\rm UV}^2}}
\label{eq15}
\end{equation}
\noindent
with the parameters
$C_{\rm IR}=1.93$, $C_{\rm UV}=1.359$,
$\omega_{\rm IR}=0.1378\,$eV, and
$\omega_{\rm UV}=13.38\,$eV.

The dielectric permittivity of ITO strongly depends on a layer
composition, thickness, etc. The literature on the subject is
quite extensive.\cite{62,63,64,65,66,67,68,69}
Specifically, the parametrization of the dielectric permittivity
of ITO was suggested\cite{69} using the Tauc-Lorentz
model\cite{70} and the Drude model.
This parametrization was used\cite{23,24} for the comparison between
the experimental data and computational results in the framework
of the Lifshitz theory. It was found, however, that the computed
magnitudes of the gradient of the Casimir force are substantially
larger than the mean measured ones.

To characterize the dielectric properties of ITO films used in our
experiment, we employed the untreated and UV-treated samples
prepared in the same way and under the same conditions as those
used in measurements of the Casimir force. The imaginary parts
of the dielectric permittivity of ITO,
${\rm Im}\,\varepsilon^{(-1)}(\omega)$,
was determined by means of ellipsometry
(J.~A.~Woollam Co.\cite{71}) for both untreated and UV-treated
samples. In the frequency region from 0.04 to 0.73\,eV the
IR-VASE ellipsometer was used.
The region of frequencies from 0.73 to 8.27\,eV was covered
with the help of VUV-VASE ellipsometer.
The experimental data obtained from ellipsometric measurements were
analyzed taking into account that the ITO resistivity decreases with
depth. This results in
different profiles of the dielectric permittivity of ITO at
different depths and typically in the so-called {\it top} and
{\it bottom} ${\rm Im}\,\varepsilon^{(-1)}(\omega)$ differing
in the frequency range $\omega<0.4\,$eV.
In Figs.~16(a,b) and 16(c,d) the experimental data for
${\rm Im}\,\varepsilon^{(-1)}(\omega)$ as a function of $\omega$
are shown by the solid lines for an untreated and UV-treated
samples, respectively. In the frequency range shown in Fig.~16(a,c)
the top and bottom permittivities coincide.
 The top ${\rm Im}\,\varepsilon^{(-1)}(\omega)$, which was found to lead
to a good agreement with the measured Casimir forces
for the untreated sample, is shown in Fig.~16(b,d) by the solid lines
in the frequency region from 0.04 to 0.8\,eV on a logarithmic scale.
It was
extrapolated in the region of low frequencies $\omega<0.04\,$eV
by means of the imaginary part of the Drude dielectric
function with the parameters
$\omega_p=1.5\,$eV, $\gamma=0.128\,$eV and
$\omega_p=1.5\,$eV, $\gamma=0.132\,$eV
for an untreated and UV-treated samples, respectively
(the dashed lines).

Precise computations of the Casimir force at separations
$a\geq 60\,$nm require knowledge of dielectric properties up to
$\omega\approx 16\,$eV. Because of this, the measured data for
${\rm Im}\,\varepsilon^{(-1)}(\omega)$
in Fig.~16(a,c) shown by the solid lines were extrapolated to higher
frequencies by means of the imaginary part of an oscillator
function
\begin{equation}
{\rm Im}\,\varepsilon^{(-1)}(\omega)=
\frac{g_0\gamma_0\omega}{(\omega^2-\omega_0^2)^2+\gamma_0^2\omega^2}.
\label{eq16}
\end{equation}
\noindent
The reasonable smooth extrapolations are bounded between the
short-dashed lines in Fig.~16(a,c). For an untreated sample [Fig.~16(a)]
the upper short-dashed line is described by Eq.~(\ref{eq16}) with the
oscillator parameters
$g_0=240.54\,\mbox{eV}^2$, $\gamma_0=8.5\,$eV and $\omega_0=9.0\,$eV.
For the lower short-dashed line we get
$g_0=111.52\,\mbox{eV}^2$, $\gamma_0=4.0\,$eV and $\omega_0=8.0\,$eV.
For the UV-treated sample [Fig.~16(c)] the oscillator parameters are
$g_0=280.28\,\mbox{eV}^2$, $\gamma_0=9.2\,$eV and $\omega_0=9.8\,$eV and
$g_0=128.28\,\mbox{eV}^2$, $\gamma_0=4.5\,$eV and $\omega_0=8.8\,$eV
for the upper and lower short-dashed lines, respectively.
As can be seen in Fig.~16(a,b,c,d), there are only minor differences
in the imaginary parts of the dielectric permittivities for the untreated
and UV-treated samples (the additional small peak near
 3\,eV for the untreated sample and insignificant variations
in the oscillator structure).
Note that the imaginary part of the ITO dielectric permittivity
suggested earlier\cite{69} and used in computations of the
gradient of the Casimir force\cite{23,24} is shown by the
long-dashed line in Fig.~16(a). It differs significantly from the
dielectric permittivity of our untreated ITO sample.
At lower frequencies deviations between the two permittivities
increase due to the larger $\omega_p=1.94\,$eV used.\cite{69}

Using the measured imaginary parts of dielectric permittivity of
ITO in Fig.~16(a,b,c,d),
the dielectric permittivities along the imaginary frequency
axis were obtained by means of the Kramers-Kronig relation.
The obtained results for the untreated sample are shown in
Fig.~17(a) by the two solids lines corresponding to the two
short-dashed lines in Fig.~16(a). In the same figure, the two dashed
lines indicate the range of dielectric permittivities along
the imaginary frequency axis for the case when the contribution of
free charge carriers were disregarded. For the UV-treated sample
the respective results obtained from the measured data in
Fig.~16(c,d) by means of the Kramers-Kronig relation are shown
in Fig.~17(b) by the two dashed lines. For the case when the
charge carriers in the UV-treated sample are disregarded, the
range of dielectric permittivities is indicated by the two
solid lines. {}From the comparison of Fig.~17(a) with Fig.~17(b)
it follows that the UV-treatment does not lead to any significant
changes in dielectric permittivity of an ITO sample as
a function of imaginary frequency.

\subsection{Theoretical results using different approaches to
the description of charge carriers}

Using the dielectric permittivities discussed in Sec.~VA
we have calculated the Casimir force $F(a,T)$ from Eq.~(\ref{eq11})
acting between an Au-coated sphere and both untreated and
UV-treated ITO samples over the range of separations from 60 to 300\,nm.
Then the surface roughness of Au and ITO films was taken into account
by means of geometrical averaging.\cite{6,10}
Note that this approximate method leads to the same results as a more
fundamental calculation based on the scattering approach\cite{72}
at short separation distances where the roughness correction reaches
maximum values (2.2\% at $a=60\,$nm and less than 1\% and 0.5\% at
$a\geq 90\,$nm and $a\geq 116\,$nm, respectively).
At separations of about the correlation length of surface roughness the
scattering approach predicts larger roughness corrections than the
method of geometrical averaging. At such large separations, however,
the effect of roughness is negligibly small and can be
disregarded.\cite{6}
As a result, the theoretical Casimir force between the rough surfaces
of an Au sphere and ITO plate was computed according to the following
expression:
\begin{eqnarray}
&&
F^{\rm theor}(a,T)=\sum_{i=1}^{N_1}\sum_{k=1}^{N_2}
v_i^{\rm (ITO)}v_k^{\rm (Au)}
\nonumber \\
&&~~~~~~~~~~~~~~~~\times
F(a+H_0^{\rm (ITO)}+H_0^{\rm (Au)}-h_i^{(\rm ITO)}
-h_k^{(\rm Au)},T),
\label{eq17}
\end{eqnarray}
\noindent
where all notations were introduced in Sec.~IIB.

The computational results using Eq.~(\ref{eq17}) in comparison with
the experimental data are shown in Fig.~18(a,b).
The theoretical Casimir forces between an Au sphere and an untreated ITO
sample are shown by the two solid lines in Fig.~18(a) over the separation
region from $a=60$ to 300\,nm. In the inset the same lines over a more
narrow separation region from 60 to 100\,nm are presented.
Computations were performed by Eqs.~(\ref{eq11}) and (\ref{eq17})
with the dielectric permittivity of Au along the imaginary frequency
axis indicated in Sec.~VA and dielectric permittivity of an untreated
ITO shown by the two solid lines in Fig.~17(a). In so doing the charge
carriers of ITO were taken into account. The experimental data for
the Casimir force (the first measurement set) are shown as crosses.
The arms of the crosses indicate the total experimental errors in the
separation distances and forces determined at a 95\% confidence level
(see Sec.~IVB). As can be seen in Fig.~18(a), the experimental data
are in a very good agreement with the theory within the limits of
theoretical uncertainties shown by the band between the two solid lines.

In Fig.~18(b) the comparison between the theoretical results
(the band between the two solid lines) and the experimental data
(crosses) is presented for a UV-treated sample. Here, to achieve the
agreement between experiment and theory, the charge carriers in the
ITO sample were disregarded. This means that the dielectric permittivity
of the UV-treated ITO shown by the two solid lines in Fig.~17(b) has been
used in computations. The inset in Fig.~18(b) demonstrates the
agreement achieved over a narrower separation region from 60 to 100\,nm.
The use of the dielectric permittivity of an ITO film with
the contribution of charge carriers disregarded
may seem somewhat unjustified
because the electric properties of an untreated and a UV-treated ITO
samples are very close.
To analyze this problem in more detail, in Fig.~19(a) we present the
comparison between experiment and theory for an untreated (the lower
pair of solid lines) and a UV-treated (the upper pair of solid lines)
samples over a separation range from 60 to 200\,nm.
As above, the experimental data are shown as crosses.
The same dielectric permittivities as in Fig.~18(a,b) were used in
computations. {}From Fig.~19(a) it is observed that theoretical results
with included (the lower pair of solid lines) and disregarded
(the upper pair of solid lines) contribution of charge carriers
do not overlap and are in very good agreement with the measurement
data for respective ITO samples.

Furthermore, in Fig.~19(b) we plot as crosses the measured Casimir
forces between a sphere and a UV-treated sample. In the same figure,
the two dotted lines show the computational results by using the
seemingly most natural dielectric permittivity of a UV-treated sample
shown by the two dashed lines in Fig.~17(b), i.e., taking into account
the free charge carriers. As can be observed in Fig.~19(b), substitution
of actual dielectric properties of a UV-treated ITO film at room
temperature in the Lifshitz theory results in drastic contradiction
with the measured Casimir forces. Note that the use of the bottom
dielectric permittivity of an ITO film discussed above instead of
the top would lead to larger in magnitude Casimir forces, i.e.,
to further increasing disagreement between experiment and theory
(note that the difference between the bottom and top permittivities
is relevant only to the contribution of free charge carriers).

In order to appreciate why the Lifshitz theory with the dielectric
permittivity disregarding charge carriers leads to
agreement with the measurement data for the UV-treated sample,
we consider the phenomenological prescription\cite{6,10} formulated
earlier to account for the
results of several experiments\cite{27,28,29,30,35,36}
discussed in Sec.~I.
 According to this
prescription, for dielectrics and semiconductors of
dielectric type free charge carriers should be disregarded,
whereas for metals they should be taken into account by
means of the plasma model. An important point in support of this
prescription is that the inclusion of relaxation properties of
electrons for metals with perfect crystal lattices and dc
conductivity for dielectrics in the Lifshitz theory results
in violation of the Nernst heat theorem.\cite{6,10}
The phenomenological prescription\cite{6,10} gave rise
to controversial discussions in the literature and even to
attempts to modify the Lifshitz theory.\cite{6,10}
One interesting consequence of this prescription is the
possibility to obtain significantly different Casimir forces
from samples with nearly equal dielectric permittivities.
To do this, one should consider a patterned Si plate with
two sections of different doping concentrations which oscillates
in the horizontal direction below an Au sphere.\cite{37}
If doping concentrations are chosen only slightly below and
above the critical value, the halves of a Si plate will be
in dielectric and metallic states, respectively.
This would lead to significantly different Casimir forces
with almost equal dielectric permittivities along the
imaginary frequency axis.

At this point one can hypothesize that the UV treatment of the
plate results in the Mott-Anderson phase
transition of an ITO film to a dielectric
state without noticeable change of its optical properties at
room temperature.  This hypothesis is supported by the observation
that the UV treatment of ITO leads to a lower mobility of charge
carriers.\cite{73}
The hypothesis proposed could be verified in future by the
investigation of electrical properties of the UV-treated ITO films
at very low temperature. Specifically, if the UV  treatment
transforms the ITO film from metallic to dielectric state, the
electric conductivity (which is similar for an
untreated and UV-treated films at room temperature) should
vanish when the temperature vanishes.

In the above computations the low-frequency behavior of the
dielectric permittivities of both Au and an untreated ITO was
described by the Drude model. We emphasize that almost the same
computational results leading to the same measure of agreement
between experiment and theory are obtained when the free charge
carriers in Au are described by the plasma model with the plasma
frequency $\omega_p=9.0\,$eV. The same is correct for untreated
ITO, but in this case the charge carriers should be described by
the plasma model with the so-called {\it longitudinal}\,\cite{63}
$\omega_p=1.3\,$eV. The value of this parameter
is determined by the physical processes at high frequencies rather
than from the extrapolation of the optical data measured at low
frequencies to zero frequency.
Note that the use of the plasma model for the description of charge
carriers for the UV-treated sample leads to the same computational
results for the Casimir force as shown by the two dashed lines in
Fig.~19(b). Thus, the inclusion of charge carriers into the Lifshitz
theory for the UV-treated sample cannot be reconciled with the
experimental data for the Casimir force shown in Fig.~19(b) as
crosses.

In this section the comparison between experiment end theory was
made using the data from the first measurement set.
The experimental data from the second set (see Table I) were also
compared with the same theoretical approaches. The results obtained
are found indistinguishable from those presented in Figs.~18(a,b)
and 19(a,b). Because of this, we do not discuss them at
greater length.

\section{Conclusions and discussion}

In the foregoing we have described the experimental observation
of the effect of significant decrease in the magnitude of the
Casimir force between an Au sphere and an ITO plate after the
UV-treatment of the latter. The main and unexpected feature of the
observed phenomenon is that a decrease in force
from 21\% to 35\% depending
on separation distance between the sphere and the plate was achieved
with no significant change of the dielectric permittivity of
the ITO film under the UV-treatment.

Measurement of the Casimir force requires precision laboratory
techniques and extreme care in all preparation procedures and
analysis. We performed our measurements using a multimode
AFM in a high vacuum chamber. In Sec.~II we have described the
setup used and all stages of the sample preparation and characterization
including the procedure of UV-treatment.

Special attention was paid to electrostatic calibrations
which are described
in Sec.~III. In the last few years calibration of the Casimir
force measurement
setup has attracted considerable interest and even become
controversial.\cite{46,47,48,49,55}
It was claimed that anomalous dependences of the residual potential
difference and separation on contact on the separation distance
observed in several experiments cast doubts on the measurements of
the Casimir force performed to date. It was also suggested that
inasmuch electrostatic calibrations are based on a fitting procedure
there is no principal difference detween independent measurements
of the Casimir
force\cite{20,21,22,23,24,25,26,27,28,29,30,35,36,37,38,40,41,42,43,44,45}
and deriving the Casimir force by means of a fit from some much larger
measured force of hypothetical origin.\cite{31}
In this respect we would like to note that the calibration consists
in determination of the parameters of a setup using well
established physical laws (in our case of electrostatics) and involves
only well understood and precisely measured forces. Because of this,
the use of some fitting procedure in the process of calibration is not,
under any circumstances, to be regarded as an evidence in favor
of the statement that
the measurement of the Casimir force is not independent. In fact, the
calibration procedure is a part of any measurement. On the contrary,
the extraction of the Casimir force by means of the fitting procedure
from much larger force, of which the major contribution
is not measured and whose origin is not
clearly understood,\cite{31} indicates that this is not an independent
measurement.

Keeping in mind these complicated issues, in Sec.~III we have
analyzed in detail different systematic deviations arising in the calibration
process. These systematic
deviations are some biases in a measurement which
always make the measured value higher or lower than the true value.
We demonstrated that if such deviations are not taken into account and
properly addressed, this results in the anomalies described in the
literature. To the contrary, we have shown that if the systematic
deviations due to finiteness of the acquisition rate and drift of
sphere-plate separation
are measured and removed by means of introducing the
 respective corrections, one arrives at the situation with no anomalies
in accordance with the well established laws of electrostatics.

In Sec.~IV we have presented our measurement results and the analysis
of random, systematic and total experimental errors. Here the main
result of our paper is demonstrated, i.e., that the UV-treatment of an
ITO film results in significant decrease in the magnitude of the
Casimir force. The histograms presented confirm that the Gaussian
distributions of the Casimir force between an Au sphere and an
untreated and, alternatively, a UV-treated sample do not overlap giving
a strong confirmation of the effect observed. The values of the total
experimental errors determined at a 95\% confidence level bring the
final confirmation to the effect of a decrease in the magnitude of
the Casimir force under UV treatment of the ITO sample.

The comparison of the experimental results obtained with the Lifshitz
theory was performed in Sec.~V. While the experimental data for an
untreated sample are in a very good agreement with
conventional applications of the Lifshitz formula for metals
(i.e., with inclusion of free charge carrier contribution),
the comparison of the data with theory for the UV-treated sample
resulted in a puzzle. The measured data were found to be in a very
good agreement with computations if the contribution of free charge
carriers is disregarded. In contrast, the inclusion of the
contribution of free charge carriers to the dielectric permittivity
of the UV-treated ITO sample
resulted in complete disagreement between the data and the
computational results. This is really puzzling if we take into
consideration that the ellipsometry measurements performed for both
the untreated and  UV-treated ITO films did not reveal any
significant differences in the imaginary parts of their dielectric
permittivities. According to the hypothetical explanation of this
phenomenon proposed in Sec.~V, the UV treatment of the ITO film
resulted in its Mott-Anderson
phase transition from the metal to dielectric state
with no significant changes in optical and electrical properties
at room temperature. Further investigations are needed for the
confirmation or rejection of this hypothesis. Specifically, one
should investigate the physical properties of complicated physical
compounds including their interaction with zero-point and thermal
fluctuations of the electromagnetic field.

Whether the proposed theoretical explanation is correct or not,
the observed phenomenon of the decreased Casimir force after the
UV treatment of an ITO sample can find prospective applications in
nanotechnology. In comparison with the case of an Au sphere
interacting with an Au plate, the Casimir force between an Au
sphere and a UV-treated ITO plate is decreased up to 65\%.
This result is of much practical importance for problems of lubrication
and stiction in micro- and nanoelectromechanical systems where the
Casimir and van der Waals forces may lead to collapse of the moving
parts of devices to the fixed electrodes, i.e., to loss of
functionality in devices. Significant decrease in the magnitude
of the Casimir force should be helpful for the resolution of such
problems.

\section*{Acknowledgments}

This work was supported by the DARPA Grant under Contract
No.~S-000354 (equipment, A.B., R.C.-G., U.M.),  NSF Grant
No.~PHY0970161 (C.-C.C., G.L.K., V.M.M., U.M.) and DOE Grant
No.~DEF010204ER46131 (G.L.K., V.M.M., U.M.).
G.L.K.\ and V.M.M.\ were also supported by the DFG Grant
BO\ 1112/20-1.


\begin{figure*}[h]
\vspace*{-10.cm}
\centerline{\hspace*{1cm}
\includegraphics{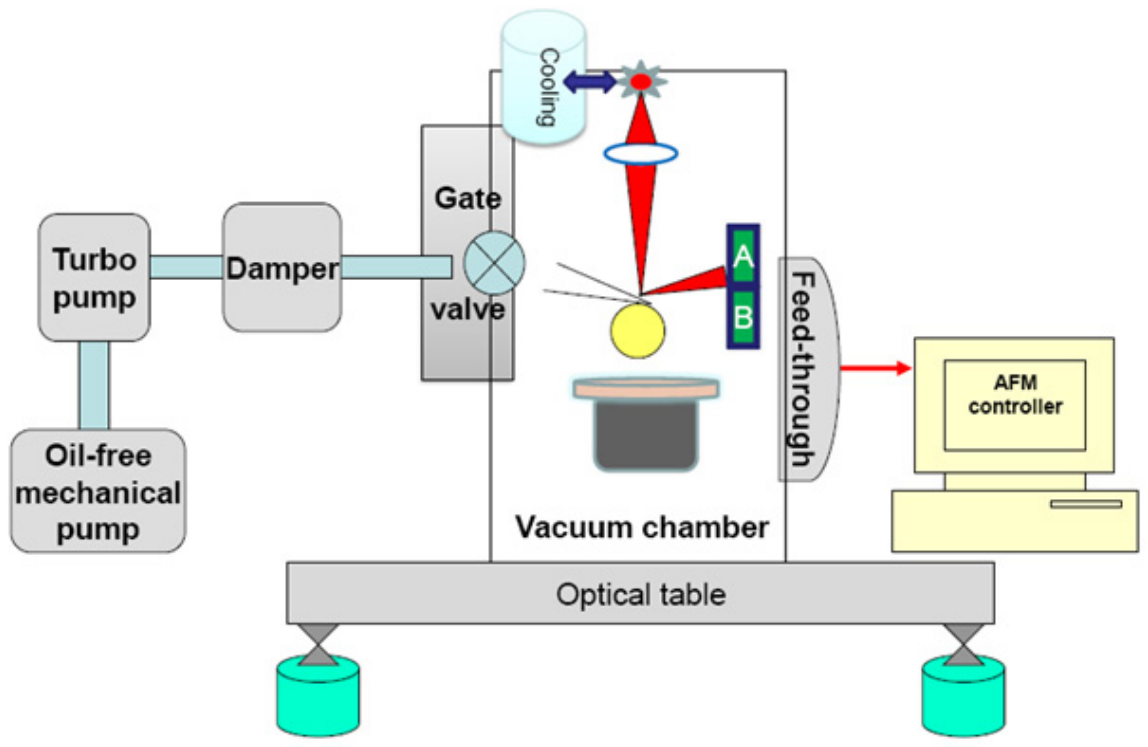}
}
\vspace*{-10.5cm}
\caption{(Color online)
Schematic of the experimental setup for measurement of
the Casimir force using an AFM (see text for further
discussion).
}
\end{figure*}
\begin{figure*}[h]
\vspace*{-8.cm}
\centerline{\hspace*{1cm}
\includegraphics{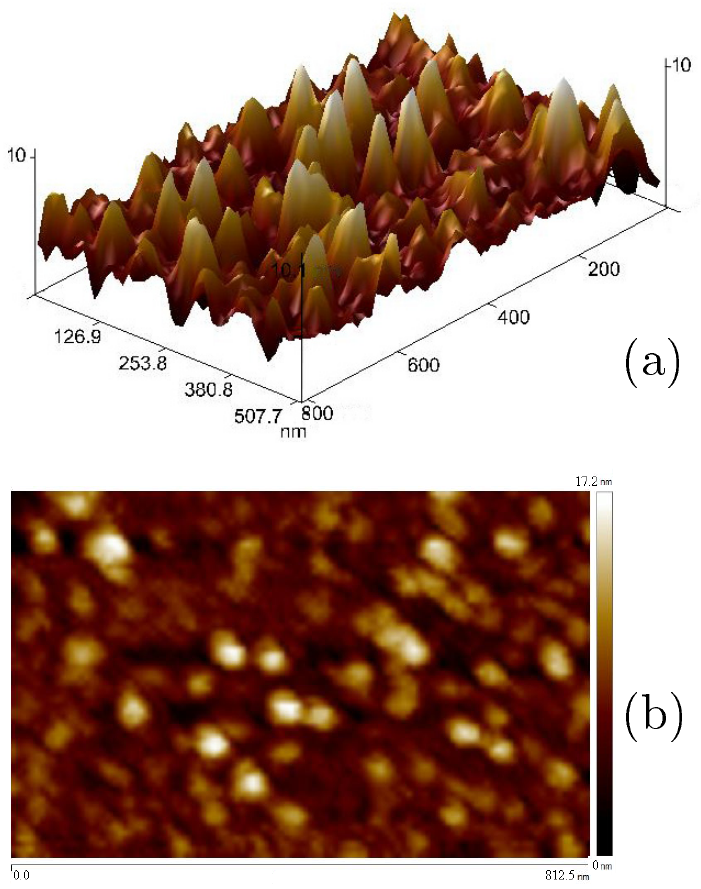}
}
\vspace*{-11.5cm}
\caption{(Color online)
(a) Typical three-dimensional AFM image of the surface of
the ITO film.
(b) Two-dimensional image of the same surface where lighter
tone corresponds to larger height.
}
\end{figure*}
\begin{figure*}[h]
\vspace*{-5.cm}
\centerline{\hspace*{1cm}
\includegraphics{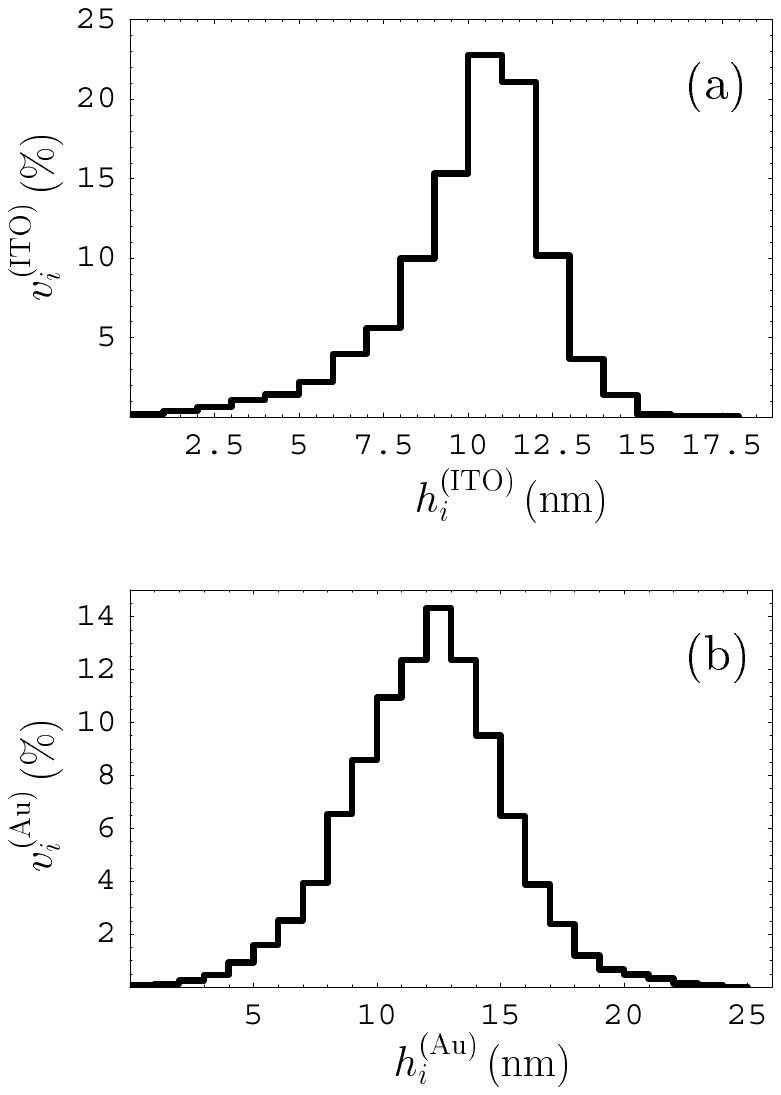}
}
\vspace*{-13.2cm}
\caption{The fractions of the area $v_i$ covered with
roughness of heights $h_i$ for (a) ITO and
(b) Au surfaces.
}
\end{figure*}
\begin{figure*}[h]
\vspace*{-8.cm}
\centerline{\hspace*{1cm}
\includegraphics{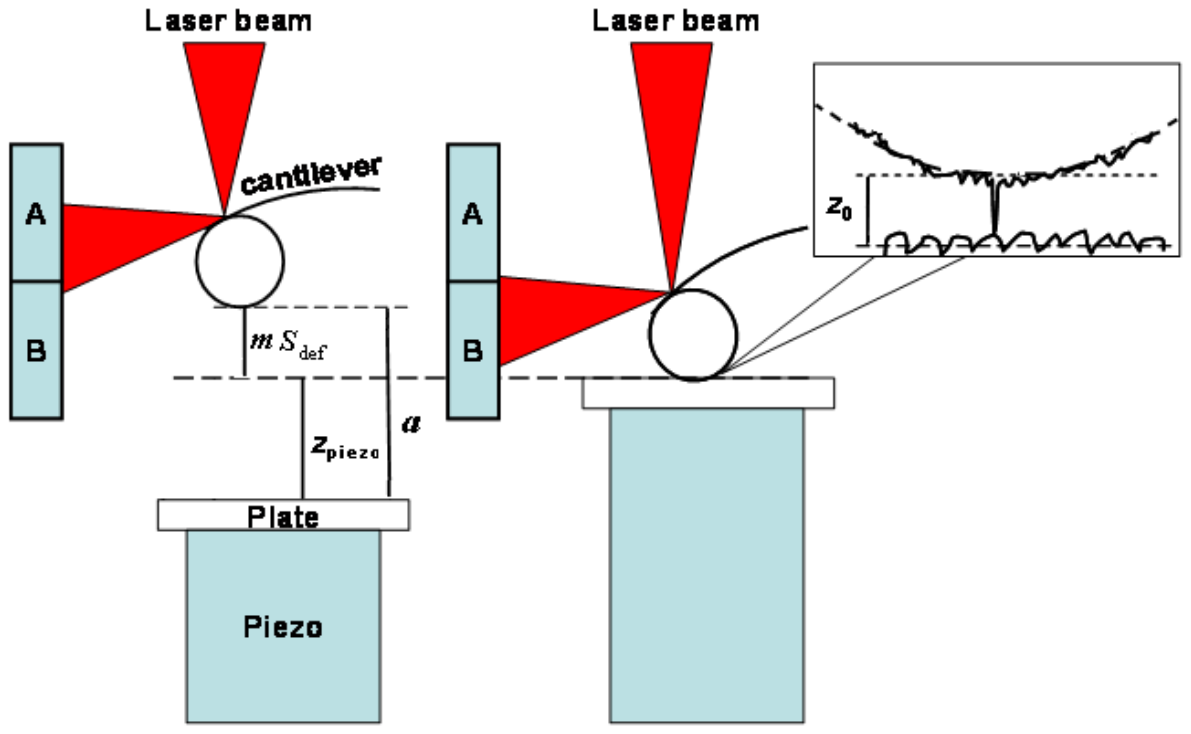}
}
\vspace*{-10.5cm}
\caption{(Color online)
Schematic explanation for the concept of the absolute
separation distance $a$, and different contributions to it,
i.e., distance traveled by the piezoelectric actuator
$z_{\rm piezo}$, distance due to the deflection of the
cantilever and separation on contact $z_0$.
}
\end{figure*}
\begin{figure*}[h]
\vspace*{-5.cm}
\centerline{\hspace*{1cm}
\includegraphics{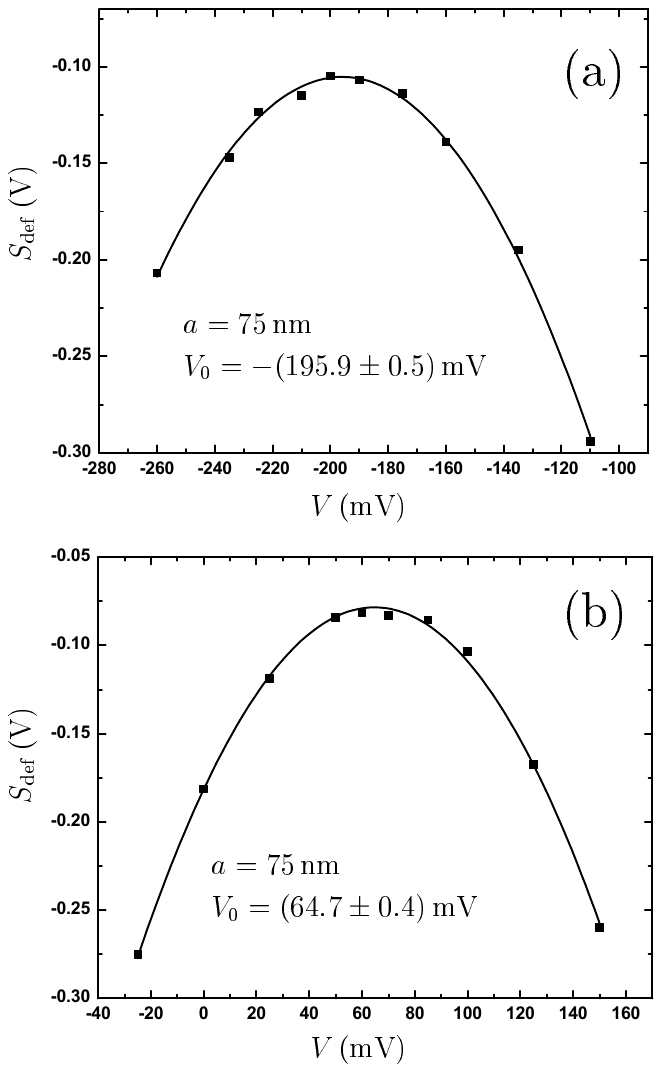}
}
\vspace*{-14.cm}
\caption{The deflection signal $S_{\rm def}$ as a function
of the applied voltage $V$ for (a) the untreated and (b)
UV-treated sample at a fixed separation $a=75\,$nm
between the sphere and the plate.
}
\end{figure*}
\begin{figure*}[h]
\vspace*{-5.cm}
\centerline{\hspace*{1cm}
\includegraphics{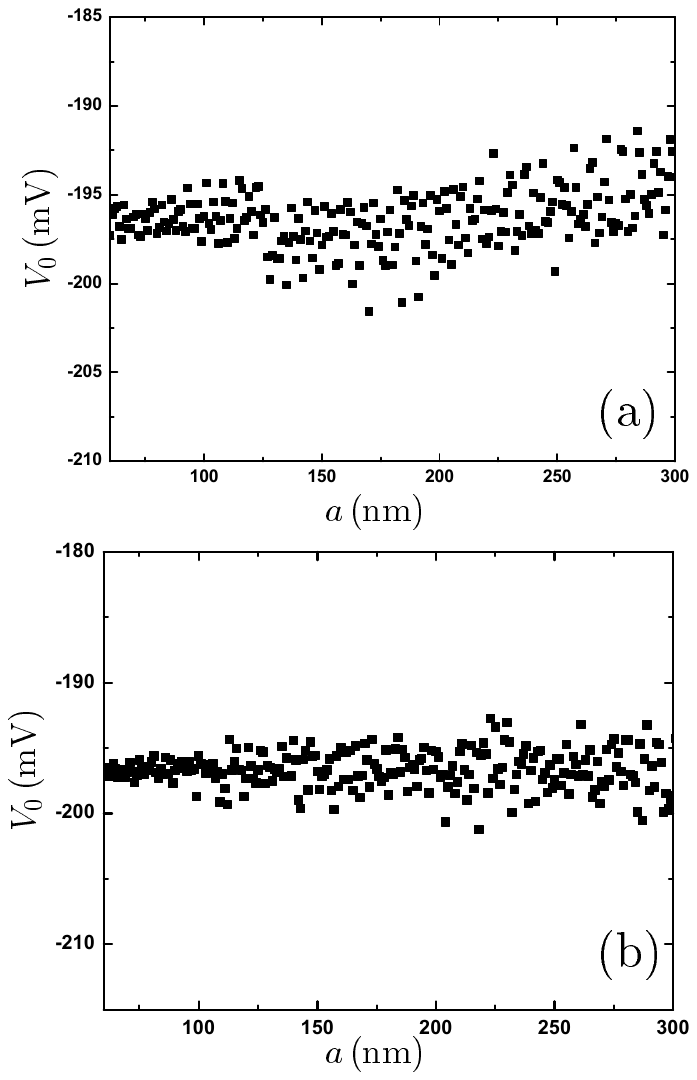}
}
\vspace*{-14.cm}
\caption{The residual potential difference $V_0$ between
the sphere and the plate surfaces as a function of
separation $a$ for the untreated sample (a) with no corrections
for systematic deviations due to drift and finite
data acquisition rate and (b) with corrections of
same deviations.
}
\end{figure*}
\begin{figure*}[h]
\vspace*{-5.cm}
\centerline{\hspace*{1cm}
\includegraphics{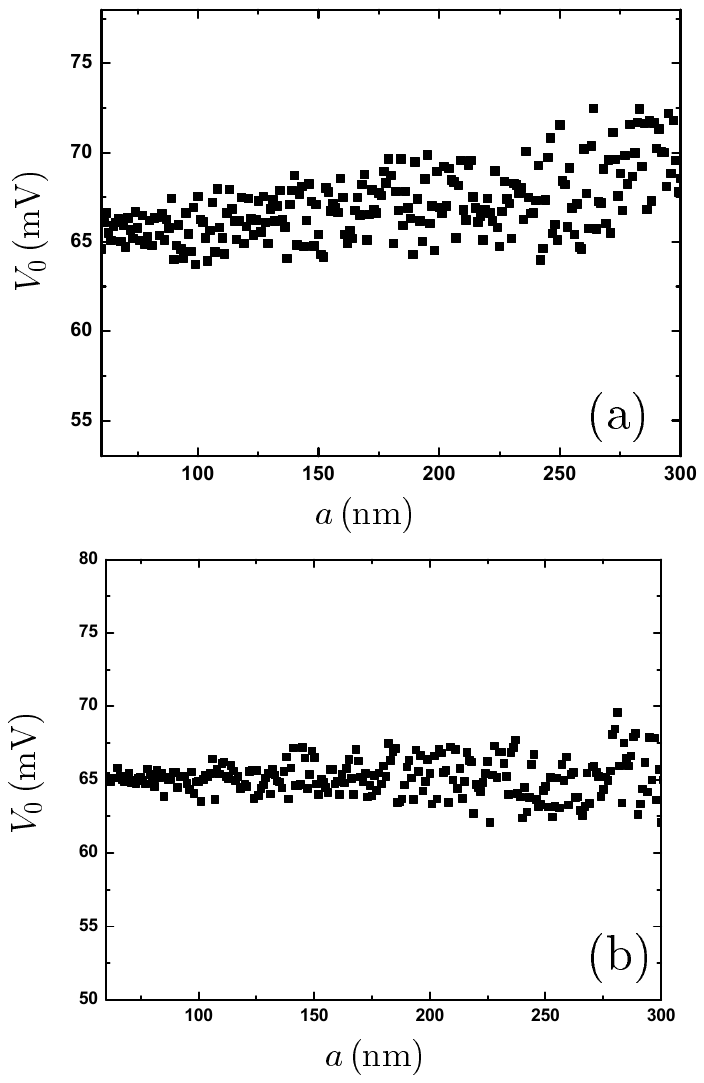}
}
\vspace*{-14.cm}
\caption{The residual potential difference $V_0$ between
the sphere and the plate surfaces as a function of
separation $a$ for the UV-treated sample (a) with no corrections
for systematic deviations due to drift and finite
data acquisition rate and (b) with corrections of
same deviations.
}
\end{figure*}
\begin{figure*}[h]
\vspace*{-5.cm}
\centerline{\hspace*{1cm}
\includegraphics{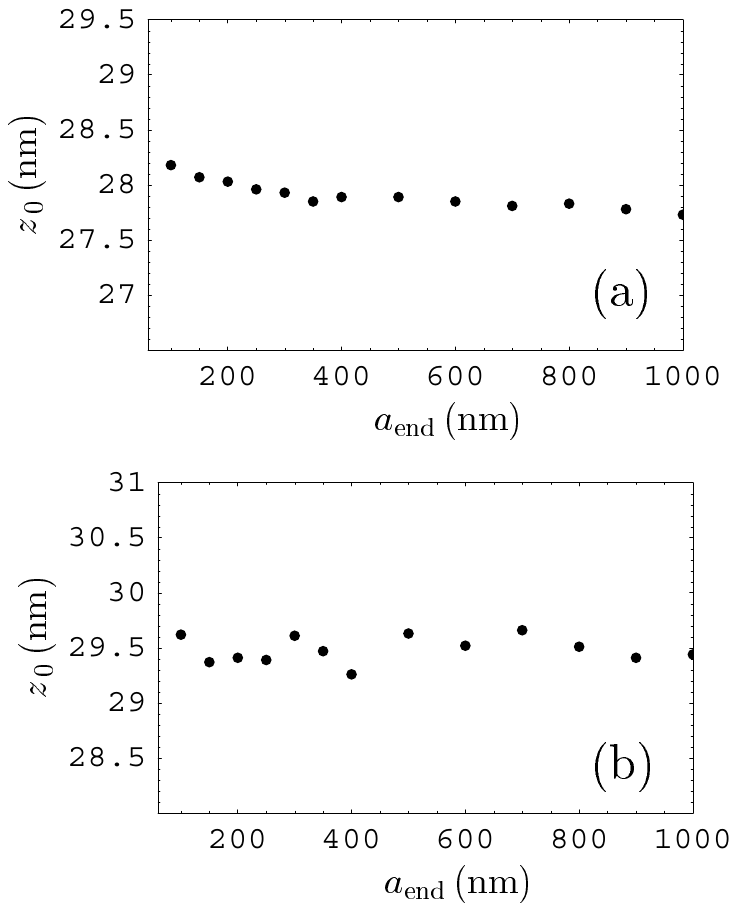}
}
\vspace*{-14.cm}
\caption{The separation on contact $z_0$ between
the sphere and the plate surfaces as a function of the end
point $a_{\rm end}$ for the untreated sample (a) with no corrections
for systematic deviations due to drift and finite
data acquisition rate and (b) with corrections of
same deviations.
}
\end{figure*}
\begin{figure*}[h]
\vspace*{-5.cm}
\centerline{\hspace*{1cm}
\includegraphics{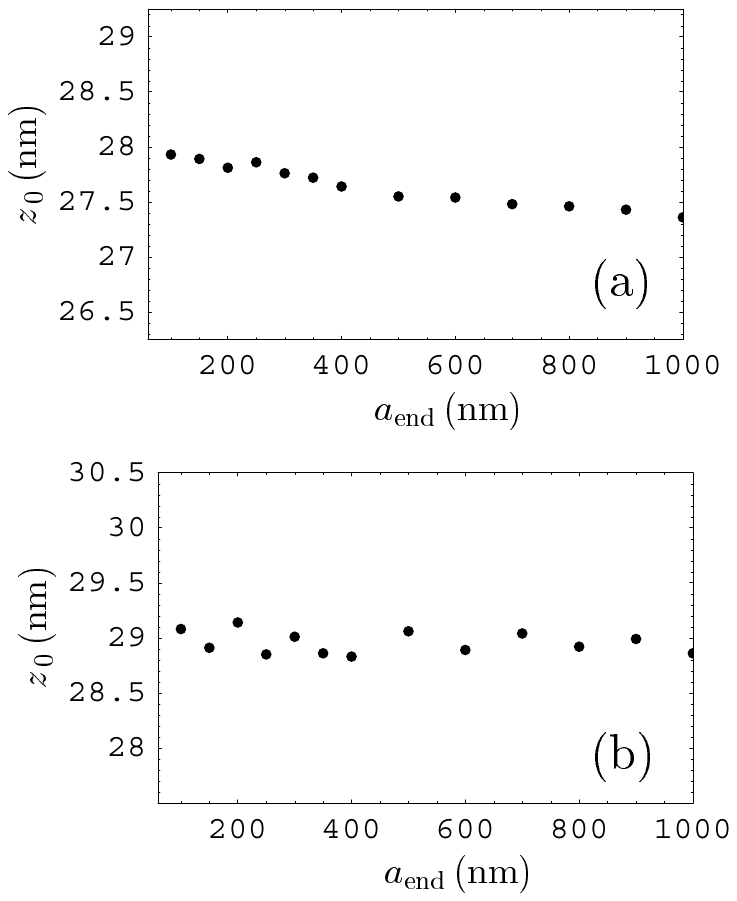}
}
\vspace*{-14.cm}
\caption{The separation on contact $z_0$ between
the sphere and the plate surfaces as a function of the end
point $a_{\rm end}$ for the UV-treated sample (a) with no corrections
for systematic deviations due to drift and finite
data acquisition rate and (b) with corrections of
same deviations.
}
\end{figure*}
\begin{figure*}[h]
\vspace*{-5.cm}
\centerline{\hspace*{1cm}
\includegraphics{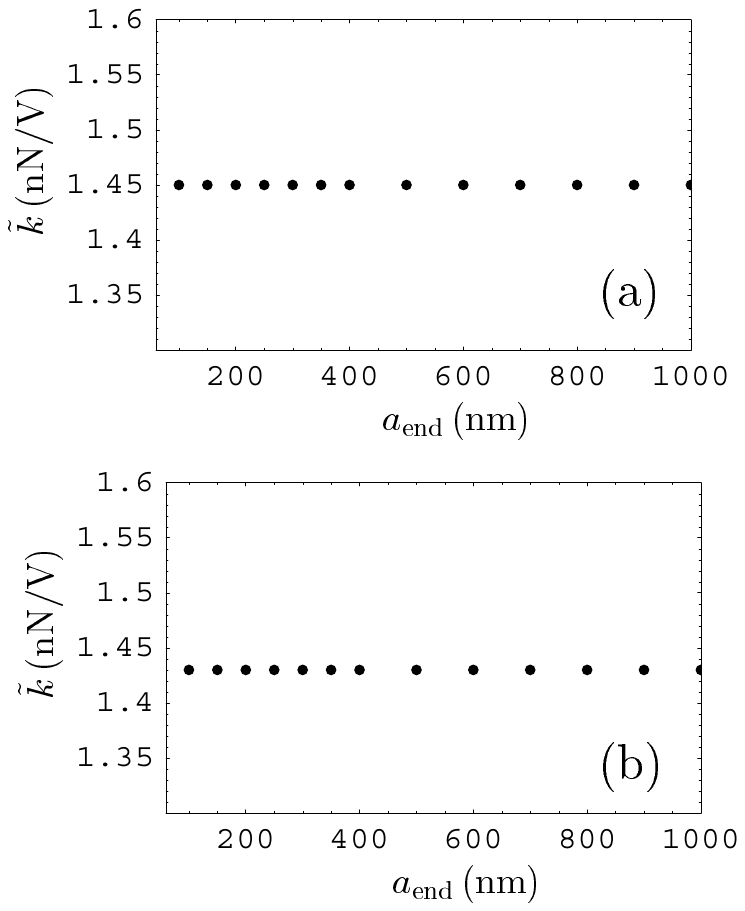}
}
\vspace*{-14.cm}
\caption{The calibration constant $\tilde{k}$
 as a function of the end
point $a_{\rm end}$ with corrections for
systematic deviations introduced
for (a) the untreated sample and (b) UV-treated sample.
}
\end{figure*}
\begin{figure*}[h]
\vspace*{-7.cm}
\centerline{\hspace*{1cm}
\includegraphics{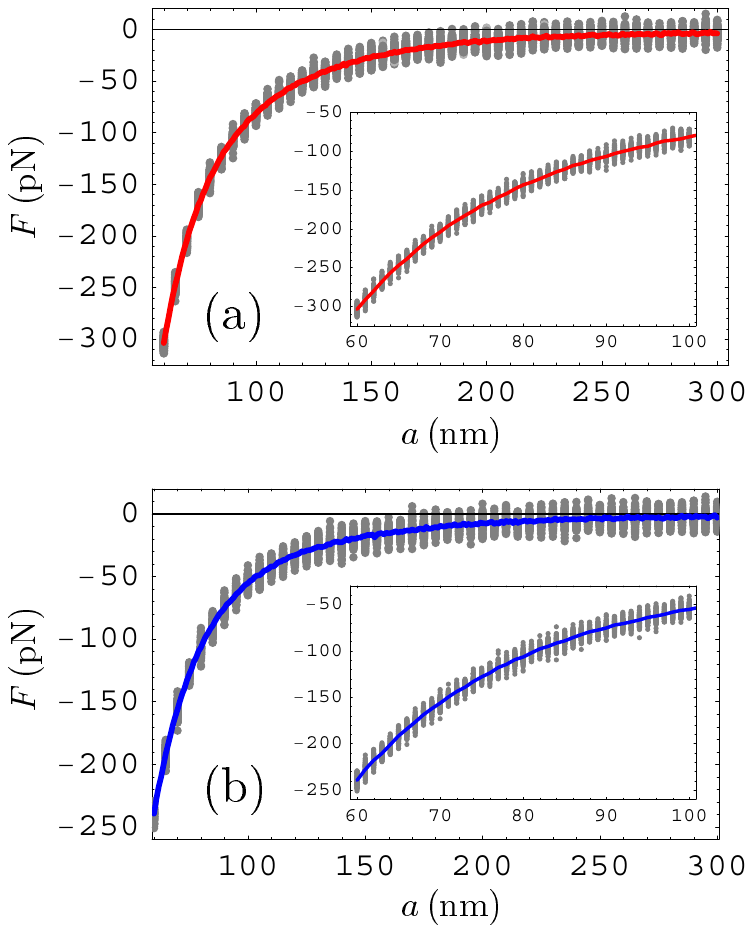}
}
\vspace*{-12.cm}
\caption{(Color online)
Mean measured Casimir forces $F$ between the sphere and
the plate as a function of separation $a$ are shown as
solid lines for (a) the untreated and (b) UV-treated sample.
In the inset the same is shown over a narrower range of
separations. All 100 individual values of the measured force
are shown as dots at separation distances at 5\,nm intervals
(1\,nm intervals in the insets).
}
\end{figure*}
\begin{figure*}[h]
\vspace*{-3.5cm}
\centerline{\hspace*{1cm}
\includegraphics{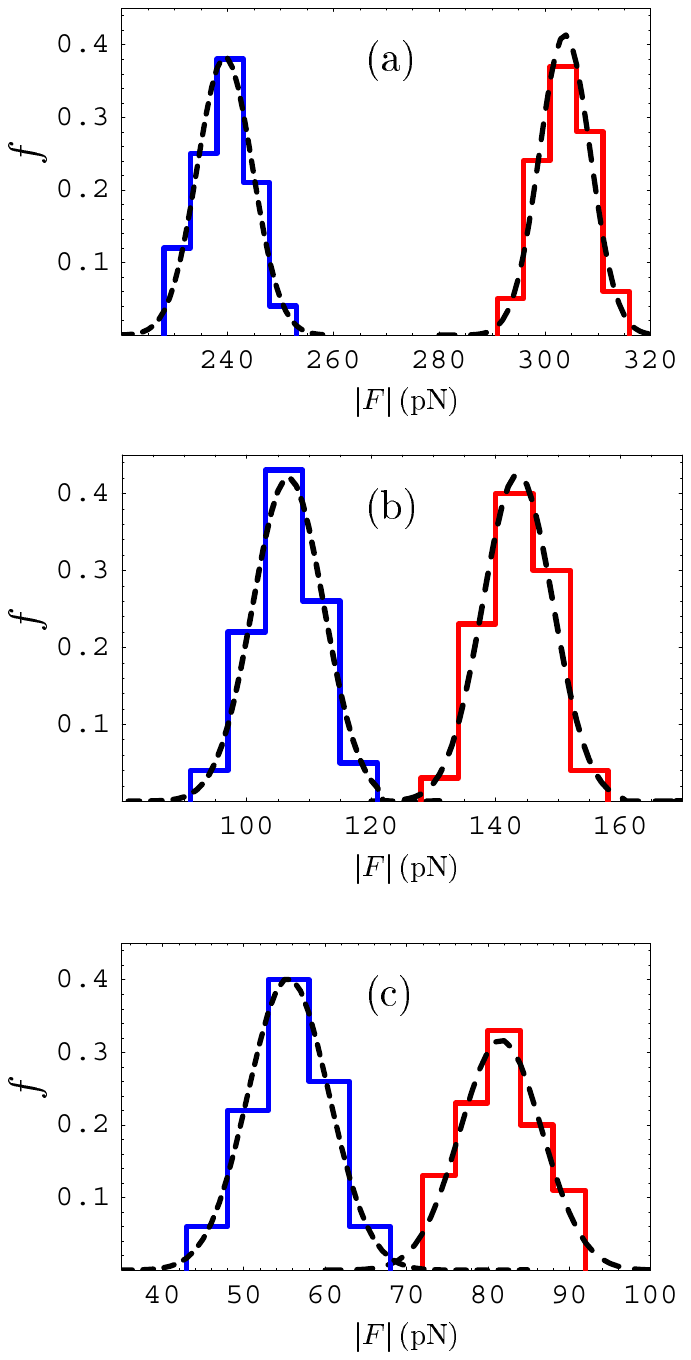}
}
\vspace*{-13.cm}
\caption{(Color online)
The histograms for measured Casimir force $F$ for the
untreated (right) and  UV-treated (left) sample at separations
(a) $a=60\,$nm, (b) $a=80\,$nm, and (c) $a=100\,$nm.
$f$ is the fraction of 100 data points having the force values
in the bin indicated by the
vertical lines. The corresponding
Gaussian distributions are shown by the
dashed lines.
}
\end{figure*}
\begin{figure*}[h]
\vspace*{-7.cm}
\centerline{\hspace*{1cm}
\includegraphics{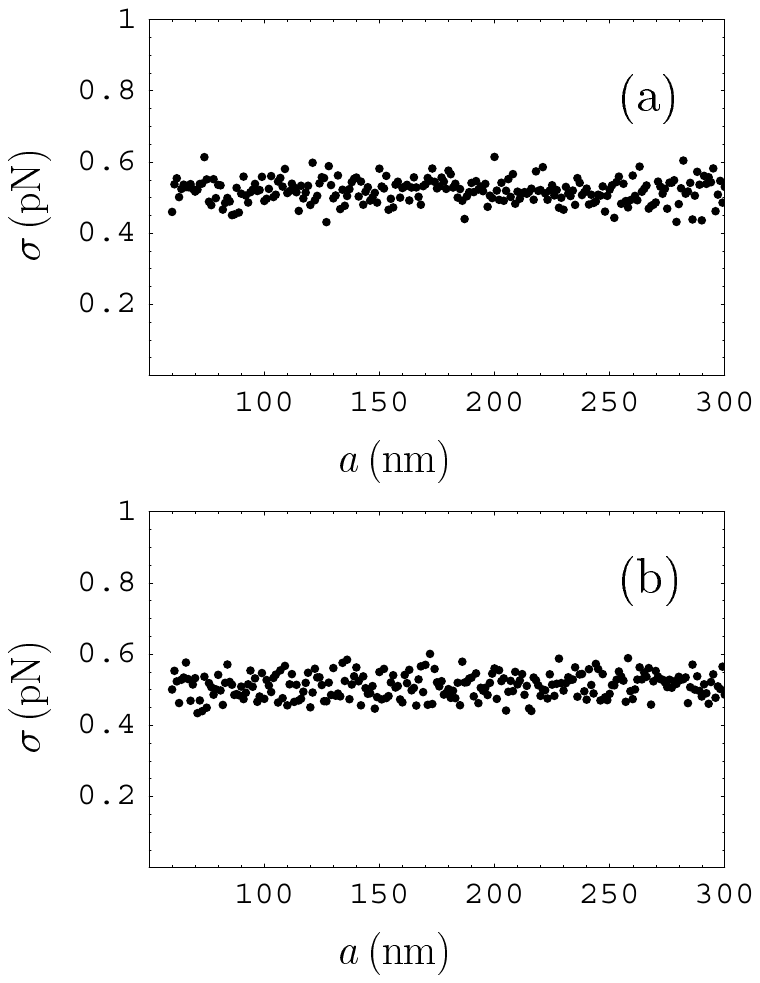}
}
\vspace*{-11.5cm}
\caption{The variance $\sigma$ of the mean Casimir force
calculated from 100 measurement results as a function of
separation $a$ for (a) the untreated and (b) UV-treated
sample.
}
\end{figure*}
\begin{figure*}[h]
\vspace*{-7.cm}
\centerline{\hspace*{1cm}
\includegraphics{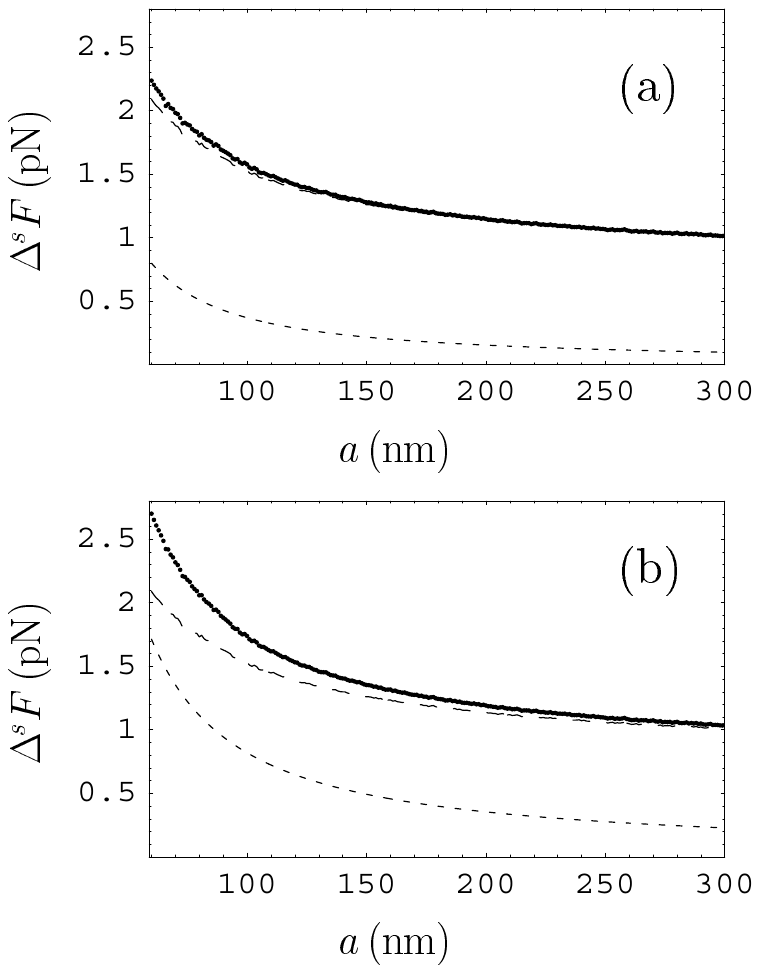}
}
\vspace*{-11.5cm}
\caption{The systematic error in the total measured force
$\Delta^{\! s}F_{\rm tot}$, the mean systematic error in
the electric force $\Delta^{\! s}F_{\rm el}$ averaged over
10 applied voltages, and the systematic error in the
Casimir force $\Delta^{\! s}F$  as a function of
separation $a$ are shown by the long-dashed lines,
short-dashed lines, and solid lines, respectively,
for (a) the untreated and (b) UV-treated
sample.
}
\end{figure*}
\begin{figure*}[h]
\vspace*{-7.cm}
\centerline{\hspace*{1cm}
\includegraphics{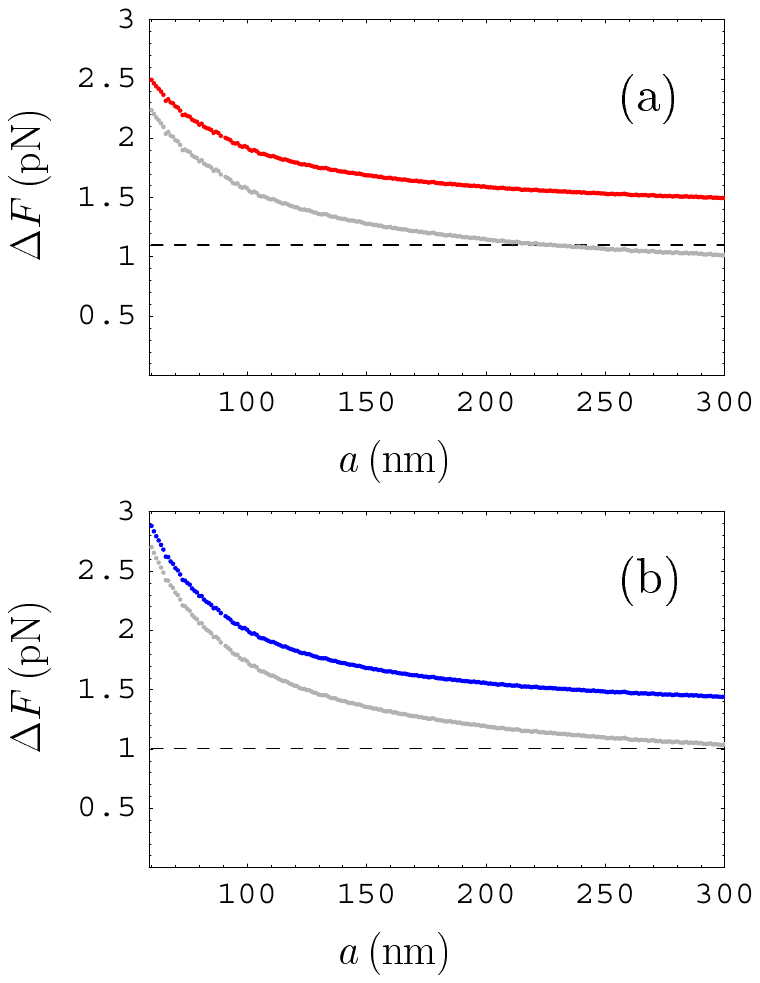}
}
\vspace*{-11.5cm}
\caption{(Color online)
The random $\Delta^{\! r}F$, systematic $\Delta^{\! s}F$,
and total $\Delta^{\! \rm tot}F$ errors in the measured
Casimir force determined at a 95\% confidence level are
shown as functions of
separation $a$ by the dashed, lower solid and upper solid
lines, respectively,
for (a) the untreated and (b) UV-treated
sample.
}
\end{figure*}
\begin{figure*}[h]
\vspace*{-7.cm}
\centerline{\hspace*{1cm}
\includegraphics{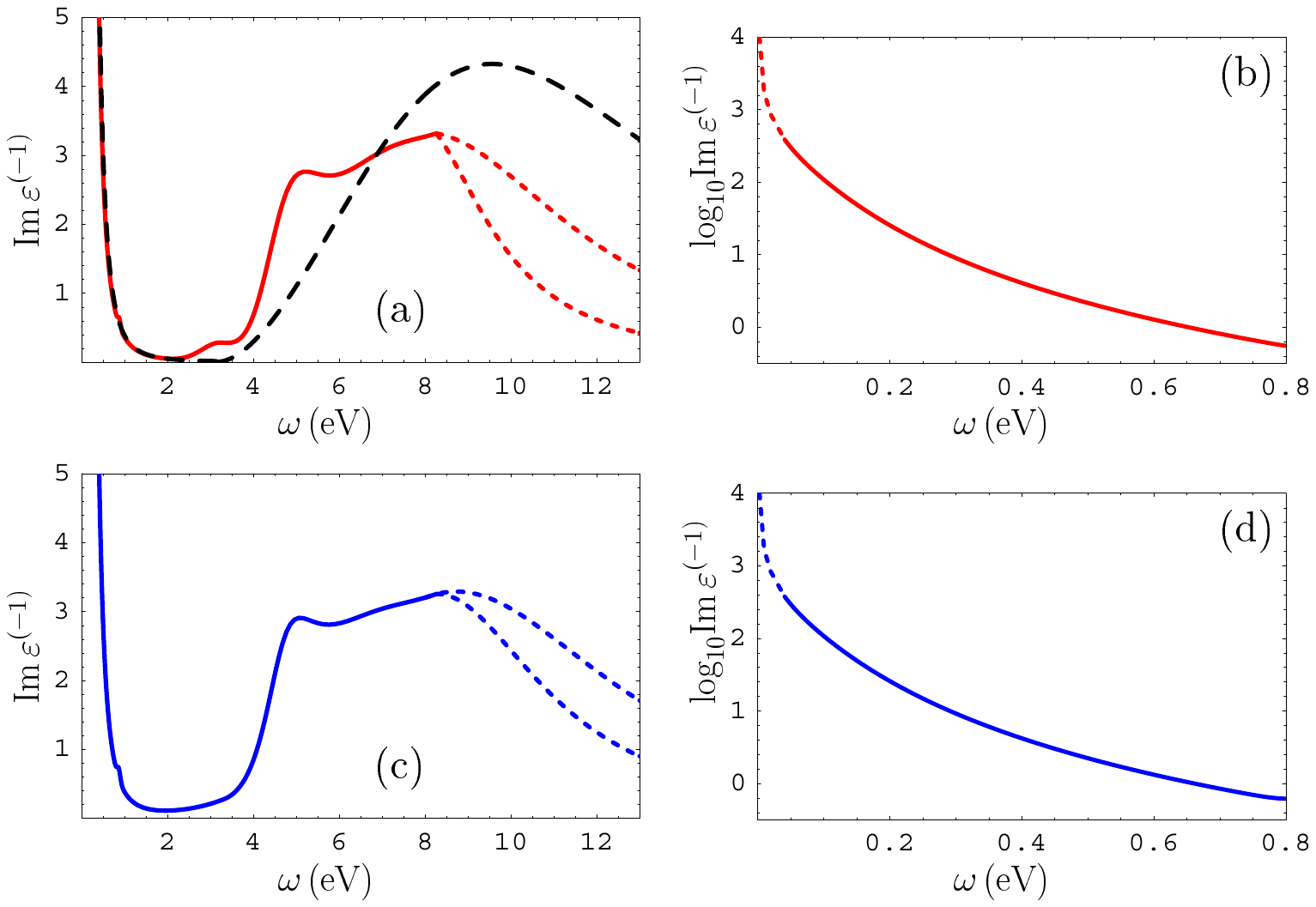}
}
\vspace*{-11.5cm}
\caption{(Color online)
The imaginary parts of dielectric permittivity of an
ITO film ${\rm Im}\,\varepsilon^{(-1)}$ obtained from
ellipsometry are shown as functions of frequency
$\omega$ with the solid lines
for (a,b) the untreated and (c,d) UV-treated sample
in different frequency regions.
The short-dashed lines (a,c) present possible extrapolations
of the data to higher frequencies (see text for further
discussion). The long-dashed line presents
${\rm Im}\,\varepsilon^{(-1)}$ from the paper
by Fujiwara and Konde\cite{69} for the untreated
ITO sample.
The dashed lines (b,d) show the extrapolation to lower
frequencies by means of the Drude model.
}
\end{figure*}
\begin{figure*}[h]
\vspace*{-7.cm}
\centerline{\hspace*{1cm}
\includegraphics{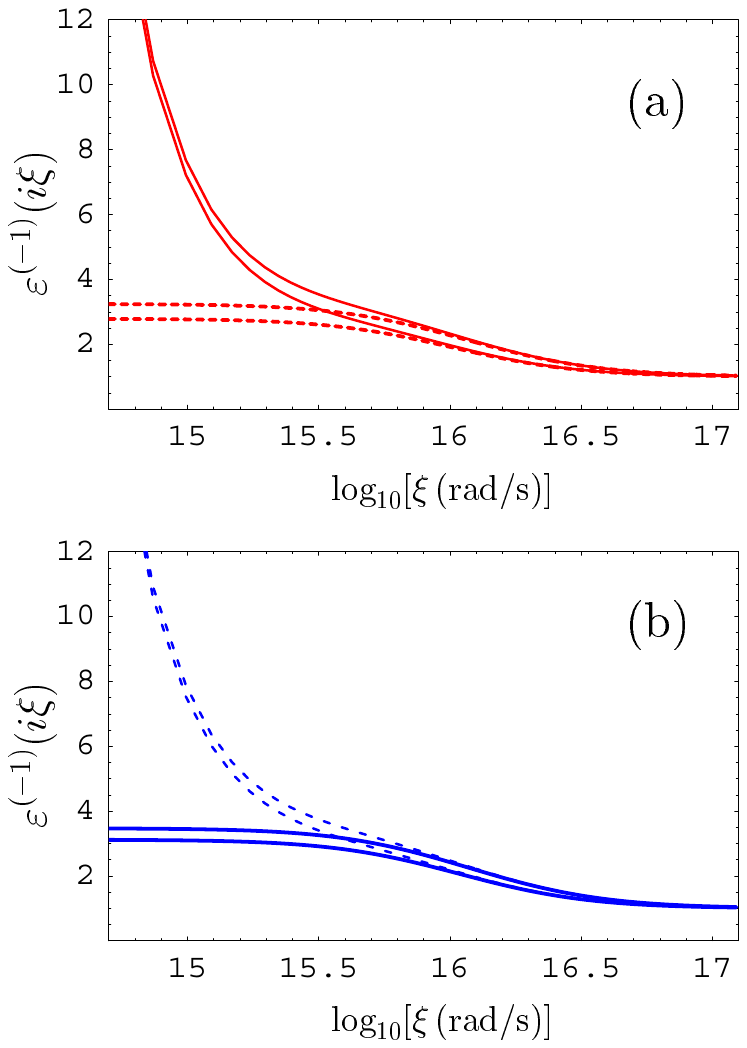}
}
\vspace*{-11.5cm}
\caption{(Color online)
The dielectric permittivity of an
ITO film $\varepsilon^{(-1)}$ as a function of
the imaginary frequency $i\xi$
for (a) the untreated and (b) UV-treated sample.
The two solid and two dashed lines are obtained with
different extrapolations of the ellipsometry data to
higher frequencies [see Fig.~16(a,b)].
In Fig.~17(a) the solid and dashed lines correspond
to included
and omitted contribution of free charge carriers,
respectively. In Fig.~17(b) the free charge carriers are
included for the pair of dashed lines and omitted for
the pair of solid lines.
}
\end{figure*}
\begin{figure*}[h]
\vspace*{-7.cm}
\centerline{\hspace*{1cm}
\includegraphics{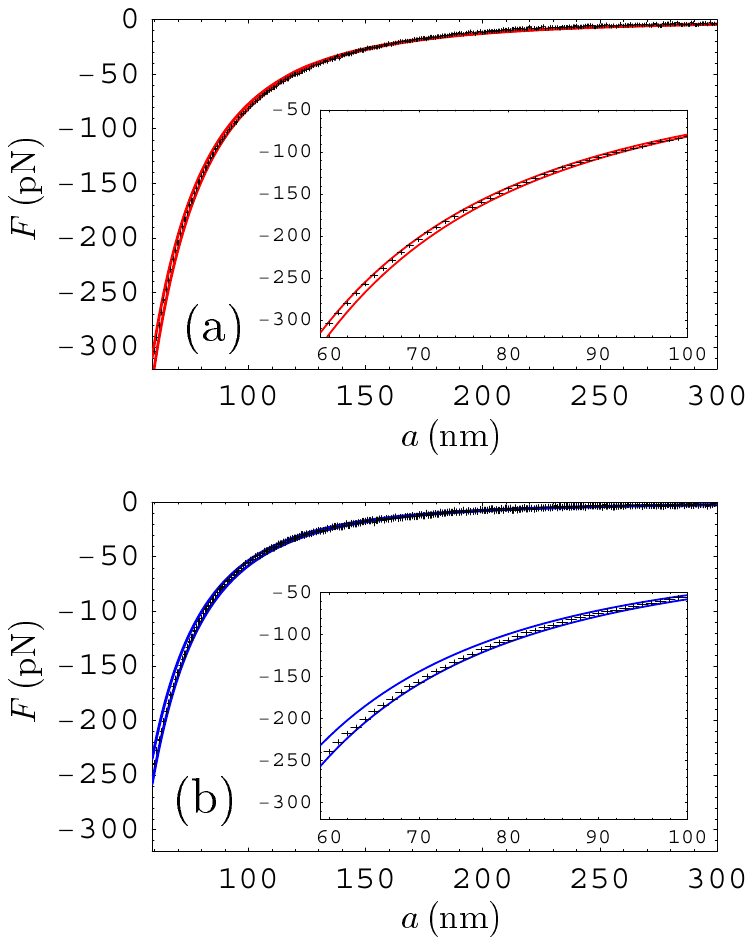}
}
\vspace*{-11.5cm}
\caption{(Color online)
The mean measured Casimir force $F$ indicated as crosses
corresponding to error bars at 95\% confidence level
and
the theoretical Casimir force $F^{\rm theor}$ shown by the
pairs of solid lines as functions of separation $a$
for (a) the untreated sample (contribution of free charge
carriers is included) and (b) UV-treated sample
(contribution of free charge carriers is omitted).
In the insets the same is shown over a narrower separation
region from 60 to 100\,nm.
}
\end{figure*}
\begin{figure*}[h]
\vspace*{-7.cm}
\centerline{\hspace*{1cm}
\includegraphics{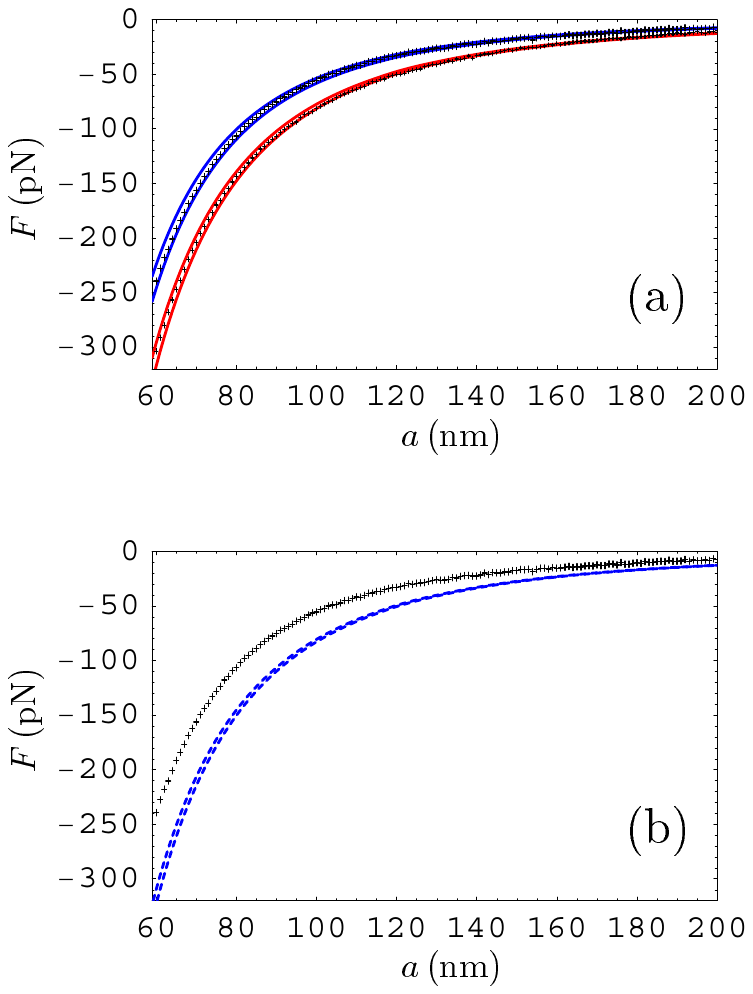}
}
\vspace*{-11.5cm}
\caption{(Color online)
(a) The mean measured Casimir force $F$ as a function of
separation $a$ is indicated as upper and lower sets of
crosses
corresponding to error bars at 95\% confidence level
for the UV-treated and untreated samples, respectively.
The respective upper and lower pairs of the solid lines show the
theoretical results computed with omitted and included
contribution of free charge carriers.
(b) The mean measured Casimir force $F$ as a function of
separation $a$ for a UV-treated sample is indicated as crosses
corresponding to error bars at 95\% confidence level.
The two dashed lines show the theoretical results $F^{\rm theor}$
computed with the contribution of free charge carriers included
for the UV-treated sample.
}
\end{figure*}
\begingroup
\squeezetable
\begin{table}
\caption{The magnitudes of the mean measured Casimir forces between
an Au sphere and an ITO plate at different separations (column 1)
for the untreated (columns 2 and 3 for the two measurement sets)
and UV-treated (columns 5 and 6 for the two measurement sets)
samples.
Columns 4 and 7 contain the total experimental errors determined
at a 95\% confidence level for the untreated and UV-treated
sample, respectively.}
\begin{ruledtabular}
\begin{tabular}{crrcrrc}
$a$&\multicolumn{3}{c}{$|F|$(pN), untreated sample}&
\multicolumn{3}{c}{$|F|$(pN), UV-treated sample}
\\
\cline{2-4}\cline{5-7}
(nm)& 1st set & 2nd set &$\Delta^{\!\rm tot}F$ &
1st set & 2nd set &$\Delta^{\!\rm tot}F$\\
\hline
60 &303.8&304.4&2.5&239.5&238.8&2.9
\\
70 &204.4&204.0&2.3&156.4&155.6&2.5
\\
80 &143.6&143.7&2.1&106.7&105.5&2.3
\\
90 &107.0&106.2&2.0&75.4&74.6&2.1
\\
100 &81.6&80.7&1.9&55.5&54.9&2.0
\\
120 &50.1&51.1&1.8&33.0&32.9&1.8
\\
140 &32.9&33.4&1.7&22.6&21.2&1.7
\\
160 &21.8&23.3&1.7&15.4&15.1&1.6
\\
180 &16.3&15.3&1.6&10.5&10.9&1.6
\\
200 &11.9&11.0&1.6&6.6&8.0&1.6
\\
220 &6.7&7.6&1.6&5.5&6.3&1.5
\\
240 &5.8&5.5&1.5&4.4&4.2&1.5
\\
260 &5.7&5.3&1.5&3.7&3.8&1.5
\\
280 &4.6&4.2&1.5&3.1&3.2&1.5
\\
300 &4.0&4.1&1.5&3.0&2.4&1.5
\end{tabular}
\end{ruledtabular}
\end{table}
\endgroup

\begin{thebibliography}{99}
\bibitem{1}
P.~W.~Milonni, {\it The Quantum Vacuum} (Academic Press, San
Diego, 1994).
\bibitem{2}
M.~Krech,
{\it The Casimir Effect in Critical Systems}
(World Scientific, Singapore, 1994).
\bibitem{3}
V.~M.~Mostepanenko and N.~N.~Trunov, {\it The Casimir Effect and
its Applications} (Clarendon, Oxford, 1997).
\bibitem{4}
K.~A.~Milton, {\it The Casimir Effect} (World Scientific,
Singapore, 2001).
\bibitem{5}
V.~A.~Parsegian,
{\it Van der Waals Forces: A Handbook for Biologists,
Chemists, Engineers, and Physicists}
(Cambridge University Press, Cambridge, 2005).
\bibitem{6}
M.~Bordag, G.~L.~Klimchitskaya, U.\ Mohideen, and
V.\ M.\ Mostepanenko, {\it Advances in the Casimir Effect}
(Oxford University Press, Oxford, 2009).
\bibitem{7}
M.~Bordag, U.~Mohideen, and V.~M.~Mostepanenko, { Phys. Rep.} {\bf
353}, 1 (2001).
\bibitem{8}
K.~A.~Milton,
J. Phys. A: Math. Gen. {\bf 37}, R209 (2004).
\bibitem{9}
S.~K.~Lamoreaux, Rep. Progr. Phys. {\bf 68}, 201 (2005).
\bibitem{10}
G.~L.~Klimchitskaya, U. Mohideen, and V.\ M.\ Mostepanenko,
 Rev. Mod. Phys. {\bf 81}, 1827 (2009).
\bibitem{11}
G.~L.~Klimchitskaya, U. Mohideen, and V.\ M.\ Mostepanenko,
 Int. J. Mod. Phys. B {\bf 25}, 171 (2011).
\bibitem{12}
A.~W.~Rodriguez, F.~Capasso, and S.~G.~Johnson,
Nature Photon. {\bf 5}, 211 (2011).
\bibitem {13}
H.~B.~G.~Casimir,
{ Proc. K. Ned. Akad. Wet. B}
{\bf 51}, 793 (1948).
\bibitem {14}
H.~B.~G.~Casimir and D.~Polder,
Phys. Rev. {\bf 73}, 360 (1948).
\bibitem{15}
H.~B.~Chan, V.~A.~Aksyuk, R.~N.~Kleiman, D.~J.~Bishop, and F.~Capasso,
Science {\bf 291}, 1941 (2001).
\bibitem{16}
H.~B.~Chan, V.~A.~Aksyuk, R.~N.~Kleiman, D.~J.~Bishop, and F.~Capasso,
Phys. Rev. Lett. {\bf 87}, 211801 (2001).
\bibitem{17}
E.~Buks and M.~L.~Roukes,
Phys. Rev. B {\bf 63}, 033402 (2001).
\bibitem{18}
E.~Buks and M.~L.~Roukes,
Europhys. Lett. {\bf 54}, 220 (2001).
\bibitem{19}
E.~M.~Lifshitz,
Zh. Eksp. Teor. Fiz. {\bf 29}, 94 (1956)
[Sov. Phys. JETP  {\bf 2}, 73 (1956)].
\bibitem{20}
F.~Chen, U.\ Mohideen, G.~L.~Klimchitskaya, and V.\ M.\ Mostepanenko,
Phys. Rev. A  {\bf 72}, 020101(R) (2005).
\bibitem{21}
F.~Chen, U.\ Mohideen, G.~L.~Klimchitskaya, and V.\ M.\ Mostepanenko,
Phys. Rev. A  {\bf 74}, 022103 (2006).
\bibitem{22}
F.~Chen, G.~L.~Klimchitskaya, V.\ M.\ Mostepanenko, and U.\ Mohideen,
Phys. Rev. Lett. {\bf 97}, 170402 (2006).
\bibitem{23}
S.~de~Man, K.~Heeck, R.~J.~Wijngaarden, and D.~Iannuzzi,
Phys. Rev. Lett. {\bf 103}, 040402 (2009).
\bibitem{24}
S.~de~Man, K.~Heeck and D.~Iannuzzi,
Phys. Rev. A {\bf 82}, 062512 (2010).
\bibitem{25}
G.~Torricelli, P.~J.~van~Zwol, O.\ Shpak, C.\ Binns,
G.\ Palasantzas, B.\ J.\ Kooi, V.\ B.\ Svetovoy, and M.\ Wuttig,
Phys. Rev. A {\bf 82}, 010101(R) (2010).
\bibitem{26}
G.~Torricelli, I.~Pirozhenko, S.\ Thornton, A.\ Lambrecht,
and C.\ Binns,
Europhys. Lett. {\bf 93}, 51001 (2011).
\bibitem{27}
F.~Chen, G.~L.~Klimchitskaya, V.\ M.\ Mostepanenko, and U.\ Mohideen,
Optics Express  {\bf 15}, 4823 (2007).
\bibitem{28}
F.~Chen, G.~L.~Klimchitskaya, V.\ M.\ Mostepanenko, and U.\ Mohideen,
Phys. Rev. B  {\bf 76}, 035338 (2007).
\bibitem{29}
R.~S.~Decca, D.~L\'opez, E.~Fischbach, G.~L.~Klimchitskaya,
 D.~E.~Krause, and V.~M.~Mostepanenko,
Phys. Rev. D {\bf 75}, 077101 (2007).
\bibitem{30}
R.~S.~Decca, D.~L\'opez, E.~Fischbach, G.~L.~Klimchitskaya,
 D.~E.~Krause, and V.~M.~Mostepanenko,
Eur. Phys. J. C {\bf 51}, 963 (2007).
\bibitem{31}
 A.~O.~Sushkov, W.~J.~Kim, D.\ A.\ R.\ Dalvit,
and S.\ K.\ Lamoreaux,
Nature Phys. {\bf 7}, 230 (2001).
\bibitem{32}
M.~Masuda and M.~Sasaki,
Phys. Rev. Lett. {\bf 102}, 171101  (2009).
\bibitem{33}
V.~B.~Bezerra, G.~L.~Klimchitskaya, U.~Mohideen, V.~M.~Mostepanenko,
and C.~Romero,
{Phys. Rev. B} {\bf 83}, 075417 (2011).
\bibitem{34}
G.~L.~Klimchitskaya, M.~Bordag,  E.~Fischbach,
 D.~E.~Krause, and V.~M.~Mostepanenko,
 Int. J. Mod. Phys. A {\bf  26}, 3918 (2011).
\bibitem{35}
J.~M.~Obrecht, R.~J.~Wild, M.~Antezza, L.~P.~Pitaevskii,
S.~Stringari, and E.~A.~Cornell,
Phys. Rev. Lett. {\bf 98}, 063201 (2007).
\bibitem{36}
G.~L.~Klimchitskaya  and V.~M.~Mostepanenko,
J. Phys. A: Math. Theor. {\bf 41}, 312002(F) (2008).

\bibitem{38}
C.-C.~Chang, A.~A.~Banishev,
G.~L.~Klimchitskaya, V.\ M.\ Mostepanenko, and U.\ Mohideen,
Phys. Rev. Lett.  {\bf 107}, 090403 (2011).
\bibitem{39}
V.~B.~Bezerra, G.~L.~Klimchitskaya, and V.~M.~Mostepanenko,
{Phys. Rev. A} {\bf 66}, 062112 (2002).
\bibitem{40}
U.~Mohideen and A.~Roy, { Phys. Rev. Lett.} {\bf 81}, 4549 (1998).
\bibitem{41}
B.~W.~Harris, F.~Chen, and U.~Mohideen, Phys. Rev. A {\bf 62},
052109 (2000).
\bibitem{42}
G.~Jourdan. A.~Lambrecht, F.~Comin, and J.~Chevrier,
Europhys. Lett. {\bf 85}, 31001 (2009).
\bibitem{43}
P.~J.~van~Zwol, V.~B.~Svetovoy, and G.~Palasantzas,
Phys. Rev. B {\bf 80}, 235401 (2009).
\bibitem{44}
H.-C.\ Chiu,  G.~L.~Klimchitskaya, V.\ N.\ Marachevsky,
V.\ M.\ Mos\-te\-pa\-nen\-ko, and U.~Mohideen,
Phys. Rev. B {\bf 80}, 121402(R) (2009).
\bibitem{45}
H.-C.\ Chiu,  G.~L.~Klimchitskaya, V.\ N.\ Marachevsky,
V.\ M.\ Mos\-te\-pa\-nen\-ko, and U.~Mohideen,
Phys. Rev. B {\bf 81}, 115417 (2010).
\bibitem{46}
W.~J.~Kim, M.~Brown-Hayes, D.\ A.\ R.\ Dalvit,
J.\ H.\ Brownell,  and R.\ Onofrio,
Phys. Rev. A {\bf 78}, 020101(R) (2008).
\bibitem{47}
R.~S.~Decca, E.~Fischbach, G.~L.~Klimchitskaya, D.~E.~Krause,
D.~L\'opez, U.\ Mohideen,  and V.\ M.\ Mostepanenko,
Phys. Rev. A {\bf 79}, 026101 (2009).
\bibitem{48}
W.~J.~Kim, M.~Brown-Hayes, D.\ A.\ R.\ Dalvit,
J.\ H.\ Brownell,  and R.\ Onofrio,
Phys. Rev. A {\bf 79}, 026102 (2009).
\bibitem{49}
R.~S.~Decca, E.~Fischbach, G.~L.~Klimchitskaya, D.~E.~Krause,
D.~L\'opez, U.\ Mohideen,  and V.\ M.\ Mostepanenko,
Int. J. Mod. Phys. A {\bf 26}, 3930 (2011).
\bibitem{50}
F.~Chen and U.~Mohideen,
Rev. Sci. Instrum. {\bf 72}, 3100 (2001).
\bibitem{51}
H.~E.~Grecco and O.~E.~Martinez,
Appl. Opt. {\bf 41}, 6646 (2002).
\bibitem{52}
W.~R.~Smythe,
{\it Electrostatics and Electrodynamics}
(McGraw-Hill, New York, 1950).
\bibitem{55}
W.~J.~Kim, M.~Brown-Hayes, D.\ A.\ R.\ Dalvit,
J.\ H.\ Brownell,  and R.\ Onofrio,
J. Phys.: Conf. Ser. {\bf 161}, 012004 (2009).
\bibitem{53}
H.-C.~Chiu, C.-C.~Chang, R.~Castillo-Garza,
F.~Chen, and U.~Mohideen,
J. Phys. A {\bf 41}, 164022 (2008).
\bibitem{54}
A.~A.~Banishev,  C.-C.~Chang, and U.~Mohideen,
Int. J. Mod.  Phys. A {\bf 26}, 3900 (2011).

\bibitem{57}
B.~Geyer, G.~L.~Klimchitskaya, and V.~M.~Mostepanenko,
Phys. Rev. A  {\bf 82}, 032513 (2010).
\bibitem{58}
M.~S.~Toma\v{s},
Phys. Rev. A  {\bf 66}, 052103 (2002).
\bibitem {59}
{\it Handbook of Optical Constants of Solids},
ed. E.~D.~Palik (Academic, New York, 1985).
\bibitem {60}
G.~Bimonte,
Phys. Rev. A {\bf 83}, 042109 (2011).
\bibitem{61}
L.~Bergstr\"{o}m,
Adv. Coll. Interface Sci. {\bf 70}, 125 (1997).
\bibitem{62}
I.~Hamberg, C.~G.~Granqvist, K.-F.~Bergren,
B.~E.~Sernelius, and L.~Engstr\"{o}m,
Vacuum {\bf 35}, 207 (1985).
\bibitem{63}
I.~Hamberg and C.~G.~Granqvist,
J. Appl. Phys. {\bf 60}, R123 (1986).
\bibitem{64}
S.~H.~Brewer and S.~Franzen,
J. Phys. Chem. B {\bf 106}, 12986 (2002).
\bibitem{65}
P.~K.~Biswas, A.~De, N.~C.~Pramanik, P.~K.~Chakraborty,
K.~Ortner, V.~Hock, and S.\ Korder,
Mater. Lett. {\bf 57}, 2326 (2003).
\bibitem{66}
J.~Ederth, P.~Johnsson, G.~A.~Niklasson, A.~Hoel,
A.~Hult{\aa}ker, P.~Heszler, and C.~G.~Granqvist,
Phys. Rev. B {\bf 68}, 155410 (2003).
\bibitem{67}
F.~Matino, L.~Persano, V.~Arima, D.~Pisignano, R.~I.~R.~Blyth,
R.~Cingolani, and R.~Rinaldi,
Phys. Rev. B {\bf 72}, 085437 (2005).
\bibitem{68}
S.~H.~Brewer, D.~Wicaksana, J.-P.~Maria, A.~I.~Kingon,
and S.~Franzen,
Chem. Phys. {\bf 313}, 25 (2005).
\bibitem{69}
H.~Fujiwara and M.~Kondo,
Phys. Rev. B {\bf 71}, 075109 (2005).
\bibitem{70}
G.~E.~Jellison, Jr. and F.~A.~Modine,
Appl. Phys. Lett. {\bf 69}, 371 (1996).
\bibitem{71}
http://www.jawoollam.com
\bibitem{72}
P.~A.~Maia Neto, A.~Lambrecht, and S.~Reynaud,
Phys. Rev. A {\bf 72}, 012115 (2005).
\bibitem{37}
R.~Castillo-Garza, C.-C.~Chang, D.~Jimenez,
G.~L.~Klimchitskaya, V.\ M.\ Mostepanenko, and U.\ Mohideen,
Phys. Rev. A  {\bf 75}, 062114 (2007).
\bibitem{73}
C.~N.~Li, A.~B.~Djuri\v{s}i\'{c}, C.~Y.~Kwong,
P.~T.~Lai, W.~K.~Chan, and S.~Y.~Liu,
Appl. Phys. A {\bf 80}, 301 (2005).

\end{thebibliography}
\end{document}